\documentclass[10pt]{article}
\usepackage{amsmath}
\usepackage{amsthm}
\usepackage{amsfonts}
\usepackage{amssymb}
\usepackage{graphics}
\usepackage{graphicx}
\usepackage{amsbsy}
\newtheorem{remark}{Remark}[section]
\newtheorem{theorem}{Theorem}[section]

\newtheorem{lemma}{Lemma}[section]
\newcommand{\LL}{{\mathcal L}}
\newcommand{\KK}{{\mathcal K}}
\newcommand{\OO}{{\mathcal O}}
\newcommand{\T}{{\mathbb T}}
\newcommand{\R}{{\mathbb R  }}
\newcommand{\pd}[2]{\frac{\partial #1}{\partial #2}}

\newcommand{\pddd}[3]{\frac{\partial^2 #1}{\partial {#2} \partial{#3}}}

\newcommand{\brk}[1]{\left( #1 \right)}

\newcommand{\cA}{\mathcal A}
\newcommand{\cL}{\mathcal L}
\newcommand{\cLo}{\cL^{OU}}
\newcommand{\rou}{\rho^{OU}}
\newcommand{\pit}{\hat{\pi}}
\newcommand{\piz}{\pi_0}
\newcommand{\eps}{\epsilon}

\begin{document}
\setlength{\baselineskip}{10pt}
\title{PERIODIC HOMOGENIZATION FOR INERTIAL PARTICLES}
\author{G.A. Pavliotis and A.M. Stuart\\
        Mathematics Institute, Warwick University\\
        Coventry, CV4 7AL, England
                    }
\maketitle

\begin{abstract}
We study the problem of homogenization for inertial
particles moving in a periodic velocity field, and subject to molecular
diffusion. We show that, under
appropriate assumptions on the velocity field, the large scale,
long time behavior of the inertial particles is governed by an
effective diffusion equation for the position variable alone. To
achieve this we use a formal multiple scale expansion in the scale
parameter. This expansion relies on the hypo-ellipticity of the underlying
diffusion. An
expression for the diffusivity tensor is found and
various of its properties studied. In particular, an expansion
in terms of the non-dimensional particle relaxation time $\tau$
(the Stokes number) is shown to co-incide with the known result
for passive (non-inertial) tracers in the singular limit $\tau \to 0$.
This requires the solution of a singular perturbation problem, achieved by means of
a formal multiple scales expansion in $\tau.$ Incompressible and potential fields are
studied, as well as fields which are neither, and theoretical findings are supported by
numerical simulations.
\end{abstract}

\section{Introduction}
\label{sec:intro}
Understanding the transport properties of particles moving in fluid flows
and subject to molecular
diffusion is a problem of great theoretical and practical interest \cite{
falko_verga, kramer}. For the purposes of mathematical analysis, the fluid velocity is
assumed to have some statistical or geometrical structure
which mimics features of real fluid flows, and yet is such that the resulting
equation describing the motion of the particle is amenable to both analysis and
efficient numerical investigations. For example the velocity may be assumed steady and
periodic in space, or may be assumed to be a Gaussian random field in space--time.
This is the problem of turbulent diffusion
\cite{csanady, kramer}.

It is often the case that the velocity field of interest is active at various length
and time scales. Consequently, the equations which govern the particle motion are
very hard to analyze directly. In such cases an effective equation which governs the
behavior of the particles at long times and large scales compared to those of the
fluid velocity is sought. The derivation of such an effective equation is based on
multiscale/homogenization techniques \cite{lions}.

This problem has been studied extensively over the last thirty
years for passive tracers, i.e. massless particles. It has been
shown that, for periodic or random velocity fields with short
range correlations, the particles perform an effective
Brownian motion. The covariance matrix of this Brownian
motion -- the {\it effective diffusivity} -- is computed through the
solution of an auxiliary equation, the {\it cell problem}. In the
case of steady periodic velocity fields the cell problem is a linear
elliptic PDE with periodic coefficients. Various properties of the
effective diffusivity have been investigated. In particular, it
has been shown that the diffusivity is always enhanced (over bare
molecular diffusion) for incompressible flows \cite{papan2, kramer, papan1} and always
depleted for potential flows \cite{vergassola}. Extensions of the above
results to the case where the molecular diffusivity is modelled as colored
noise have also been analyzed \cite{castiglione_inert}.

However, there are various applications where the particles cannot be modelled as
tracers and inertial effects have to be taken into account.  The need for adequate
modelling and analysis of the motion of inertial particles in various applications
in science and engineering has been recognized over the last few years with a number
of publications in this direction \cite{eaton, falko_clouds, shaw_clouds}. As well
as bringing important physical features into the problem, the presence of inertia
also renders the mathematical study of the resulting equations more delicate. The
purpose of the present paper is to study the long time, large scale behavior of
inertial particles moving in periodic velocity fields, under the influence of
molecular diffusion. Both analytical and numerical techniques are used. From a
physical standpoint the main interest in this work is perhaps the enormous
enhancement in diffusion which can be introduced through relatively small inertial
effects; see section \ref{sec:num_exp}. From the mathematical standpoint, the main
interest is perhaps (i) the hypo-elliptic structure of the differential operators
arising in the multi-scale expansions (section \ref{sec:homog}) and (ii) the
elucidation of the passive tracer limit, from within the inertial particles
framework, again by use of multiple scales expansions (section \ref{sec:stau}).

We consider the diffusive behaviour of trajectories satisfying the equation
\begin{equation}
\tau\ddot{x}=v(x)-\dot{x}+\sigma \dot{\beta},
\label{eqn:langevin}
\end{equation}
\begin{figure}
\begin{center}
\includegraphics[width=2.9in, height = 2.9in]{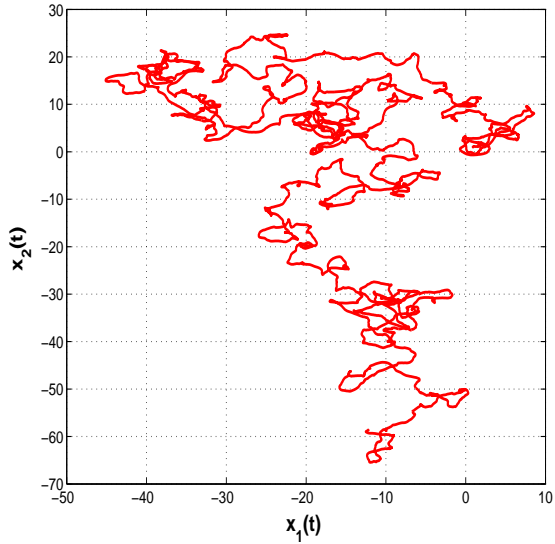}
\caption{The two components of the solution of \eqref{eqn:langevin}, \eqref{e:tg}
with $\tau = \sigma =1.0$.}
\end{center}
\label{fig:tg_traj}
\end{figure}
where $\beta(t)$ is a standard Brownian motion in $\R^d, \, d \geq 1$. In Figure
\ref{fig:tg_traj} we plot a sample trajectory of \eqref{eqn:langevin} with dimension
$d=2$, $\tau=1$ and $\sigma=1$; the velocity $v(x)$ is given by the Taylor--Green
field:
\begin{equation}
v(x) = \nabla^\bot \psi_{TG}(x), \quad \psi_{TG}(x) =  \sin(x_1) \sin(x_2).
\label{e:tg}
\end{equation}
This figure strongly suggests that the long time behavior of the particles is
diffusive. Our goal is to make this type of observation precise. 
We provide a mathematical formula for the effective diffusivity, and
study its properties by a combination of analysis and Monte Carlo
simulations. The results raise a number of interesting questions
of physical interest, regarding behaviour of the effective diffusivity,
and also provide a framework for  mathematical analysis to further
probe questions of physical interest. 

Problems of this type have already been studied in a variety of situations:
\begin{enumerate}
\item When $\sigma=0$, $\nabla \cdot U(x) \equiv 0$ and
\begin{equation}
\label{eq:velf}
v(x)=U(x)+\delta \tau U(x) \cdot \nabla U(x)
\end{equation}
we obtain the following model for the motion of a particle
in an incompressible steady velocity field \cite{maxey_1, maxey_2}:
\begin{equation}
\ddot{x} = \delta U(x)\cdot \nabla U(x) + \frac{1}{\tau}(U(x) - \dot{x}).
\label{eqn:crisanti}
\end{equation}
\begin{figure}
\begin{center}
\includegraphics[width=2.9in, height = 2.9in]{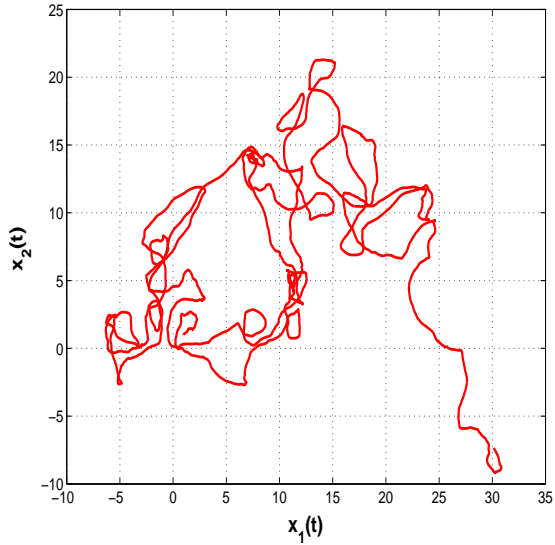}
\caption{The two components of the solution of \eqref{eqn:crisanti}, for $U(x)$ given by 
\eqref{e:tg}, with $\delta = 0.7$ and $\tau =2.0$.}
\end{center}
\label{fig:cris_traj}
\end{figure}
Here $x(t)$ denotes the particle position and $U(x)$ is the fluid velocity.  The
parameter $\delta = \frac{\rho_f}{\rho_p}$ is the ratio between the fluid density
and the particle density and $\tau$ denotes the Stokes number. Equation
\eqref{eqn:crisanti} becomes Stokes' law when $\delta = 0$, i.e. when the particles
are much denser than the surrounding fluid. The model \eqref{eqn:crisanti} was
analyzed numerically for two dimensional steady incompressible cellular flows by
Crisanti et al. \cite{crisanti_inert}. Their numerical experiments showed that for
particles slightly denser than the surrounding fluid ($1 - \delta \approx 0$ and
$\delta<1$), and for $\tau = 1$, the particles perform an effective Brownian motion
with a well defined diffusion coefficient  The term inertial diffusion was
introduced to describe this phenomenon. In Figure
2 we present a sample trajectory of \eqref{eqn:crisanti} with $U(x)$ given by the
Taylor Green velocity field \eqref{e:tg}, for $\tau =2.0$ and $\delta=0.7$. The
figure again suggests long time behavior of the particle trajectories which is
diffusive. This phenomenon has been observed and analyzed--mostly numerically--by
various authors \cite{crisanti_inert, maxey_4,  maxey_5, wang, maxey_3}. The range
of $\delta$ in which Brownian motion behaviour is observed appears to be connected
to the linear stability properties of the equilibria of \eqref{eqn:crisanti}; see
the discussion in section \ref{sec:num_exp}.

\item When $\sigma=0$, $\nabla \cdot U(x)=0$ and $v(x)=U(x)$
(i.e. $\delta=0$ in \eqref{eqn:crisanti})
a similar investigation was carried out by Wang et al. in
\cite{wang} in the case $d=3$.
They studied equation (\ref{eqn:crisanti})
numerically for the three--dimensional
ABC flow. It is well known that the streamlines of this flow are
chaotic \cite{aref_1}. It was exhibited numerically that, for
sufficiently small $\tau$, the particles perform Brownian motion.
The resulting diffusion coefficient was evaluated numerically and
its dependence on $\tau$ was investigated.
\item When $\sigma>0$ and $v(x)=-\nabla V(x)$
it is possible to study the effective behavior of particles moving in a
periodic or random potential flow, under the influence of
molecular diffusion. This has been addressed by various researchers
\cite{ garnier, lebo_einstein, papan_varadhan, rodenh}. In
this case the particle motion is governed by the Langevin
equation:
\begin{equation}
\ddot{x} = - \nabla V(x) -\dot{x} + \sigma \dot{\beta},
\label{eqn:lang_intro}
\end{equation}
where $\beta(t)$ is the standard Brownian motion in $\mathbb{R}^d$, $\sigma > 0$ and
$V(x)$ is a potential which can be either periodic or random. It was shown by
Papanicolaou and Varadhan \cite{papan_varadhan} that, for sufficiently smooth random
potentials $V(x)$, the particles perform an effective Brownian  motion, at long
length and time scales, with a nonnegative--definite effective diffusivity. A
similar result was proved for periodic potentials in \cite{rodenh} and various
refinements were analyzed in \cite{garnier, lebo_einstein}. The analysis of equation
\eqref{eqn:lang_intro} is greatly simplified by the fact that the explicit form of
the invariant distribution associated to the particle position and velocity is
known.
\item The problem of averaging for equations of the form \eqref{eqn:langevin} with
periodic coefficients was studied by Freidlin in \cite{freidlin8}. He considered the
equation
\begin{equation*}
\tau \ddot{x} = v \left(\frac{x}{\eps} \right) -
   \dot{x} + \sigma \left(\frac{x}{\eps} \right) \dot{\beta},
\end{equation*}
where $v(x), \, \sigma(x)$ are smooth, $1$--periodic functions and the matrix $a(x) =
\sigma(x) \sigma(x)^*$ is non--degenerate. It was shown in \cite{freidlin8} that as
$\tau, \, \eps$ tend to $0$, the particle position $x$ converges to the
solution of a first order SDE with constant coefficients, leading to linear
in time mean flow, with superimposed Brownian fluctuations.
%
%
The averaged coefficients of the limiting SDE depend on how fast
$\eps$ tends to $0$ relative to $\tau$ and two different cases have to be
distinguished.
\end{enumerate}
The purpose of the present paper is to study the long time,
large scale behavior of inertial particles
satisfying \eqref{eqn:langevin}
when the velocity field is periodic with
period $1$: $v(x + \hat{e}_j) = v(x), \; \; j = 1, \dots, d$ where
$\{\hat{e}_j \}_{j =1}^d$ are the unit vectors in $\mathbb{R}^d$.
We show, by applying multiscale techniques to the backward
Kolmogorov equation associated to (\ref{eqn:langevin}), that,
provided the drift term is centered with respect to the invariant distribution
of the stochastic process corresponding to \eqref{eqn:langevin}, the long
time,
large scale behavior of the inertial particles is governed by an
effective Brownian motion. The effective diffusivity tensor $\KK$ is
computed through the solution of the cell problem which is an
equation posed on $\R^d \times \T^d$, where $\T^d$ is the $d$--dimensional unit
torus. It is shown that this
tensor is nonnegative and that,
consequently, the effective dynamics is well--posed. We emphasize
that our analysis does not assume any specific structure on the
drift term and that, in particular, it is valid for both
divergence free as well as potential flows, provided that a
centering condition on $v(x)$ is satisfied. However, in the case of
potential flows, the effective diffusivity is shown to be depleted, for all
Stokes' numbers, when compared with bare molecular diffusion; this
generalizes a known result for passive tracers.

Our theoretical findings are supported by numerical simulations which
probe the dependence of the effective diffusivity on parameters such as $\tau$ and,
when the velocity field is given by \eqref{eq:velf}, $\delta.$ The equations of
motion (\ref{eqn:langevin}) are solved for the Taylor--Green flow \eqref{e:tg} and
the long time behavior of the particle trajectories is shown numerically to be that
of a Brownian motion. The numerically computed effective diffusivity is compared
with the enhancement for passive tracers, i.e. when $\tau = 0$, and it is shown that
the presence of inertia enhances the diffusivity beyond the enhancement for the
passive tracers. The problem is also analyzed through a small $\tau$ expansion of
the effective diffusivity. We show that, to leading order in $\tau$, the effective
diffusivity is equal to that arising from the homogenization of passive tracers.
We also compute the $\mathcal{O}(\tau)$ correction to the effective diffusivity for
one dimensional gradient flows of the type \eqref{eqn:lang_intro} and show that it
is always negative. Thus, in the one dimensional case and for $\tau$
sufficiently small but positive, the diffusivity is depleted over bare molecular diffusion even
further than the depletion which occurs when $\tau = 0$.

Here we analyze the problem of homogenization of inertial particles and
derive the formula for the effective diffusivity using formal multi--scale
calculations. We emphasize that the basic homogenization result derived in this
paper can be proved rigorously using techniques from stochastic analysis, in
particular the martingale central limit theorem \cite{kipnis}; we refer to
\cite{per_hom_hypoell} for the rigorous proof of the homogenization theorem.

The paper is organized as follows. In section \ref{sec:homog_first_order} we review
homogenization for passive tracers. The derivation of the homogenized equation for
the general second order stochastic differential equation \eqref{eqn:langevin} with
periodic velocity field $v(x)$ is discussed in section \ref{sec:homog}. Various
properties of the effective diffusivity are derived and analyzed in section
\ref{sec:eff_diff_prop}. The homogenization result for the equation
\eqref{eqn:crisanti}, subject to molecular diffusion, is also presented there. The
small $\tau$ expansion for $\KK$ is studied in section \ref{sec:stau}. Numerical
experiments for the Taylor--Green flow are presented in section \ref{sec:num_exp}.
Section \ref{sec:conclusions} contains our conclusions. Finally, some technical
results which are needed for the rigorous justification of the multi--scale method
employed in this paper are proved in the appendix.

\section{Homogenization for Passive Tracers}
\label{sec:homog_first_order}

In this section we review the homogenization result for passive
tracers -- i.e. massless particles -- advected by a velocity field
$v(x)$ and subject to molecular diffusion:
\begin{equation}
\dot{x} = v(x) + \sigma \dot{\beta}.
\label{eqn:smoluchowski}
\end{equation}
We do this in order to emphasize the structural similarities
between this case and the inertial case \eqref{eqn:langevin} which is the focus
of the paper.

The velocity field $v(x)$ is assumed smooth and periodic with period $1$.
We will take $\sigma > 0$. This problem has been extensively
analyzed, cf. \cite{lions, kramer, papan1,  vergassola}. Here we
merely outline the main steps in the derivation of the homogenized
equation. Notice that we do not assume that the flow is either
incompressible or potential. We also remark that we have chosen to
base our analysis on the Backward Kolmogorov equation as opposed
to Forward Kolmogorov--i.e. the Fokker--Planck-- equation, which
is more customary.  Of course, the two approaches are equivalent, 
provided that the field $v(x)$ is sufficiently smooth. We feel, however,
that the analysis based on the backward Kolmogorov equation is cleaner
than that based on the Fokker--Planck equation. Furthermore, the
backward Kolmogorov equation is the starting point for the rigorous justification of the results
reported here, see \cite{ per_hom_hypoell, kipnis}.

The process $x(t)$ whose evolution is governed by
\eqref{eqn:smoluchowski} is ergodic on the torus ${\mathbb T}^d$,
\cite[ch. 3]{lions} for details. This in particular implies the existence
of a unique, smooth, invariant density $\piz(x)$ which is the
unique solution to the stationary Fokker--Planck equation. We will
assume that the drift term averages to $0$ with respect to this
invariant density $\piz$:
\begin{equation}
\int_{\mathbb{T}^d} v(x) \piz(x) \, dx = 0.
\label{eqn:centering_1d}
\end{equation}
The above centering condition ensures that the long time, large
scale behavior of the particle is diffusive and, in particular,
excludes the possibility of the existence of an effective drift.
It is automatically satisfied for potential flows (for which the
invariant density $\piz(x)$ is the Boltzmann distribution) and it
reduces to the condition that the average of the drift over the
unit cell vanishes for incompressible flows, since in this case
$\piz(x) = 1$ \footnote{It would seem more natural from a physical
point of view to impose the condition $\int_{\T^d} v(x) \, dx =
0$, rather than condition \eqref{eqn:centering_1d}, since the
invariant density itself depends on the drift
$v(x)$. But, as we will see, condition \eqref{eqn:centering_1d}
is natural from a mathematical point of view. If, instead, the
condition $\int_{\T^d} v(x) \, dx =0$ is imposed,
then an effective drift can appear and then a
Galilean transformation with respect to this effective drift can
bring the problem to the form that we consider.}.

\subsection{The Rescaled Process}
We use the scaling property of Brownian motion $\beta(c t) =
\sqrt{c} \beta(t)$ in law \cite[Lemma 2.9.4]{karatzas} to deduce
that, under the re-scaling $t \to t/\epsilon^2$, $x \to x/\epsilon$
we obtain
\begin{equation}
\dot{x} = \frac{1}{\epsilon}
v(\frac{x}{\epsilon}) + \sigma \dot{\beta}.
\nonumber
\end{equation}
Setting $z=x/\epsilon$ gives the system of SDEs
\begin{subequations}
\begin{equation}
\dot{x} = \frac{1}{\epsilon}v(z) + \sigma \dot{\beta},
\end{equation}
\begin{equation}
\dot{z} = \frac{1}{\epsilon^2}v(z) + \frac{\sigma}{\epsilon} \dot{\beta},
\end{equation}
\label{eqn:sdexz}
\end{subequations}
with the understanding that $z \in {\mathbb T}^d$ and $x \in {\mathbb R}^d.$ This
clearly exhibits the fact that the problem possesses three time scales of
${\cal O}(1), {\cal O}(\epsilon)$ and ${\cal O}(\epsilon^2)$.
We now average out the fastest scale, given by $z$.

\subsection{Mutiscale Expansion}
The backward Kolmogorov equation for \eqref{eqn:sdexz} is
\begin{equation}
\frac{\partial u}{\partial t} = \left(
\frac{1}{\epsilon^2} \mathcal{L}_0 + \frac{1}{\epsilon}
\mathcal{L}_1 + \mathcal{L}_2 \right) u,
\label{eqn:back_kolmog_1d_resc}
\end{equation}
with
\begin{eqnarray}
\mathcal{L}_0 &=& v(z) \cdot \nabla_z + \frac{\sigma^2}{2} \Delta_z,
\nonumber \\
\mathcal{L}_1 &=& v(z) \cdot \nabla_x + \sigma^2 \nabla_z .\nabla_x,
\nonumber \\
\mathcal{L}_2 &=& \frac{\sigma^2}{2} \Delta_x. \nonumber
\end{eqnarray}
Here $\LL_0$ is a uniformly elliptic operator with periodic boundary
conditions on ${\mathbb T}^d$. We look for a solution of equation
\eqref{eqn:back_kolmog_1d_resc} which has the form:
\begin{equation}
u = u_0 + \epsilon u_1 + \epsilon^2 u_2 + \dots
\nonumber
\end{equation}
with $u_i = u_i(x, z, t), i =1,2, \dots$. Substituting the above
ansatz into (\ref{eqn:back_kolmog_1d_resc}) we obtain the following
sequence of equations:
\begin{eqnarray}
\mathcal{L}_0 \, u_0  &=&  0,
\nonumber \\
\mathcal{L}_0 \, u_1  &=& - \mathcal{L}_1 \, u_0, \label{eqn:oept} \\
\mathcal{L}_0 \, u_2 &=&  - \mathcal{L}_1 \, u_1 - \LL_2 \, u_0 +
\frac{\partial u_0}{\partial t}. \nonumber
\end{eqnarray}
The ergodicity of the process generated by $\LL_0$ implies that
the null space of the operator $\mathcal{L}_0$ consists of
functions which are constants in $z$. We let $\piz$ be the invariant
density which satisfies the equation
\begin{equation}
\label{eqn:id1}
\LL_0^* \piz(z) = 0.
\end{equation}
Thus the equation $\mathcal{L}_0 f = g$ has a solution if and only if
$g$ averages to $0$ over $\mathbb{T}^d$ with respect to $\piz(z)$:
\begin{equation}
\label{eqn:avge}
\langle g \rangle_{\piz}:= \int_{\T^d} g(z) \piz(z) \, dz = 0.
\end{equation}
We refer to  \cite[ch. 3]{lions} for proofs of these results.

In view of the discussion in the previous paragraph, from the first equation
in (\ref{eqn:oept}) we deduce that $u_0 = u_0(x,t)$. Furthermore,
the centering condition on the drift term $v(x)$ implies that
$$ \int_{\mathbb{T}^d} \mathcal{L}_1 u_0(x,t) \piz(z) \, dz =
\brk{\int_{\mathbb{T}^d} v(z) \piz(z) \, dz } \cdot \nabla_x u_0 =
0,$$ and consequently the second equation in (\ref{eqn:oept}) is well--posed.
We can solve this equation using separation of variables:
\begin{equation}
u_1(x,z,t) = \chi(z) \cdot \nabla_x u_0(x,t)
\nonumber
\end{equation}
where the corrector field $\chi(z)$ satisfies the  {\it cell
problem} :
\begin{equation}
\mathcal{L}_0 \chi(z) = -v(z).
\label{eqn:cell_1d}
\end{equation}
Notice that, in view of the centering condition on the drift term,
the cell problem is well--posed and it admits a unique solution.

Now we proceed with the final equation in (\ref{eqn:oept}). We apply the
solvability condition to obtain:
\begin{eqnarray}
0     & = & \int_{\mathbb{T}^d} \left( \mathcal{L}_1 u_1 +
            \mathcal{L}_2 u_0 - \frac{\partial u_0}{\partial t} \right)
            \piz(z) \, dz
      \nonumber \\ & = &
       - \frac{\partial u_0}{\partial t} + \frac{\sigma^2}{2}
       \Delta_x u_0 + \left( \int_{\mathbb{T}^d} \left[ v(z) \otimes \chi(z) +
       \sigma^2 \nabla_z \chi(z) \right]  \piz(z) \, dz \right) : D_x^2 u_0
       \nonumber \\ & = &
       - \frac{\partial u_0}{\partial t} + \mathcal{K}:D_x^2 u_0.
\nonumber
\end{eqnarray}
We have used the notation $D_x^2 u_0$ for the Hessian of $u_0$:
$D_x^2 u_0 = \left( \pddd{u_0}{x_i}{x_j} \right)_{i,j=1}^d$ and $A:B$ for the product
of the matrices $A$ and $B$. Moreover, $\otimes$ stands for the tensor product between
two vectors.

From the above equation we deduce that:
\begin{equation}
\frac{\partial u_0}{\partial t} = \mathcal{K}_{ij}
\pddd{u_0}{x_i}{x_j},
\label{eqn:homog_first_ord}
\end{equation}
where the summation convention has been used. The {\it effective
diffusivity} $\KK$ is
\begin{equation}
\mathcal{K}_{ij} = \frac{\sigma^2}{2} \delta_{ij} +
\int_{\mathbb{T}^d} v_j(z) \chi_i(z) \piz(z) \, dz + \sigma^2
\int_{\mathbb{T}^d} \pd{\chi_i}{z_j}(z) \piz(z) \, dz.
\label{eqn:eff_diff_1d}
\end{equation}
 Equation (\ref{eqn:homog_first_ord}) is the
backward Kolmogorov equation associated to a Brownian motion
$X(t)$ with covariance matrix $\mathcal{K}$. We remark that these
formal calculations can be justified rigorously using either
 energy estimates \cite{lions}, the method of two--scale convergence
 \cite{allaire} or probabilistic methods \cite{bhatta, pardoux}.
\subsection{Well-posedness of the Limiting Equation}
The limiting backward Kolmogorov equation
\eqref{eqn:homog_first_ord} is well--posed, i.e. the effective
diffusivity is a nonnegative matrix. To see this, we first
observe that
\begin{eqnarray}
\mathcal{L}_0^* (f \piz) & = & - \piz \mathcal{L}_0 f + \sigma^2 ( \Delta_z
f ) \piz + \sigma^2 \nabla_z f \cdot \nabla_z \piz
 \nonumber \\ & = &
-\pi_0 \mathcal{L}_0 f+\sigma^2\nabla_z \cdot \{(\nabla_z f)\pi_0\},
\label{eqn:what}
\end{eqnarray}
for every smooth periodic function $f(z).$
Consequently:
\begin{equation}
 \int_{\mathbb{T}^d} f \left( \mathcal{L}_0 f \right) \piz \, dz
      =    - \frac{\sigma^2}{2}
            \int_{\mathbb{T}^d} |\nabla_z f |^2
           \piz \, dz,
\label{eqn:gen_negative_first_ord}
\end{equation}
in view of \eqref{eqn:what} and an integration by parts. Now let
$a$ be an arbitrary vector in $\R^d$ and let $\chi_0 = a . \chi$ where $\chi $
is the solution of the cell problem (\ref{eqn:cell_1d}). The
scalar quantity $\chi_0$ satisfies the equation
\begin{equation}
\label{eqn:cell1}
\mathcal{L}_0 \chi_0 = -  a \cdot v.
\end{equation}
We consider the effective diffusivity along the
direction $a$. We use (\ref{eqn:gen_negative_first_ord}) to
obtain:
\begin{eqnarray}
\begin{array}{ccc}
a \cdot \mathcal{K}  a & = &  \frac{\sigma^2}{2} |a|^2 +
                           \int_{\mathbb{T}^d}  (a\cdot v)   \chi_0  \piz \,
                           dz + \sigma^2 \int_{\mathbb{T}^d}   (\nabla_z
                           \chi_0 \cdot a) \piz \, dz \\
& = &  \frac{\sigma^2}{2} |a|^2 -
                           \int_{\mathbb{T}^d}  ({\cal L}_0 \chi_0)\chi_0\piz \,
                           dz + \sigma^2 \int_{\mathbb{T}^d}   (\nabla_z
                           \chi_0 \cdot a) \piz \, dz \\
& = &  \frac{\sigma^2}{2} |a|^2 +\frac{\sigma^2}{2}
                           \int_{\mathbb{T}^d}  |\nabla_z \chi_0|^2 \piz \,
                           dz + \sigma^2 \int_{\mathbb{T}^d}   (\nabla_z
                           \chi_0 \cdot a) \piz \, dz \\
& = &  \frac{\sigma^2}{2} \int_{\mathbb{T}^d} |\nabla_z \chi_0 + a|^2 \piz dz.
\end{array}
\label{eqn:thisone}
\end{eqnarray}
Using the notation \eqref{eqn:avge}, the penultimate line in \eqref{eqn:thisone}
may be written as
\begin{equation}
\label{eqn:thistwo}
a\cdot {\mathcal K}a =   \frac{\sigma^2}{2} |a|^2 +
\frac{\sigma^2}{2}\langle |\nabla_z \chi_0|^2 \rangle_{\piz}+\sigma^2
\langle \nabla_z \chi_0 \cdot a \rangle_{\piz}.
\end{equation}
From \eqref{eqn:thisone} we deduce that
the effective diffusivity is indeed nonnegative and the well
posedness of the effective equation \eqref{eqn:homog_first_ord} is
demonstrated.
\subsection{Incompressible Flows}
Whether the diffusivity is enhanced or depleted depends on the
specific properties of the periodic drift term. For the case where
the flow is steady and either divergence free or potential more
detailed information can be obtained. In particular, for
incompressible flows we have that $\piz(z) = 1$. Consequently the
last integral on the right hand side of the penultimate line in
(\ref{eqn:thisone}) vanishes on account of the periodicity of
$\chi_0(z)$. Thus, the effective diffusivity along the vector $a$
becomes:
\begin{equation}
a . \mathcal{K} a = \frac{\sigma^2}{2} |a|^2 +
                          \frac{\sigma^2}{2}\int_{\T^d}
                          |\nabla_z \chi_0 |^2 \piz(z) \, dz.
\label{eqn:eff_diff_incomp_fst_ord}
\end{equation}
This shows that transport is always enhanced over bare molecular diffusion,
for incompressible flows \cite{papan1}.

\subsection{Potential Flows}
When $v(z) = - \nabla V(z)$, the invariant
density $\piz(z)$ is the Boltzmann distribution:
$$
\piz(z) = \frac{1}{Z} \exp(-\frac{2}{\sigma^2} V(z)), \; \; Z =
\int_{\T^d} \exp(-\frac{2}{\sigma^2} V(z)) \, dz.
$$
An integration by parts, together with the periodicity of
$\chi_0(z), \, \piz(z)$ and equation
(\ref{eqn:gen_negative_first_ord}), gives:
\begin{equation}
\sigma^2 \int_{\mathbb{T}^d} \nabla_z \chi_0(z) . a \piz(z) \, dz = -
\sigma^2 \int_{\T^d} |\nabla_z \chi_0(z) |^2 \piz(z) \, dz,
\nonumber
\end{equation}
and consequently, from the penultimate line in \eqref{eqn:thisone}, we find that
\begin{equation}
\begin{array}{ccc}
a . \mathcal{K}a
&=&\frac{\sigma^2}{2} |a|^2 - \frac{\sigma^2}{2} \int_{\T^d}
                          |\nabla_z \chi_0 |^2 \piz(z) \, dz,\\
&=&\frac{\sigma^2}{2} |a|^2 - \frac{\sigma^2}{2}
\langle |\nabla_z \chi_0 |^2 \rangle_{\piz}
\end{array}
\label{eq:square}
\end{equation}
with $\piz(z)$ being the Boltzmann distribution. This shows that
transport is  always depleted, compared with bare molecular diffusion,
for potential flows \cite{vergassola}.
%
%
\section{Homogenization for Inertial Particles}
\label{sec:homog}
In this section we will derive the homogenized equation which
describes the motion of inertial particles at long times and large
scales using multi-scale techniques. The equation of motion for
the inertial particles is \eqref{eqn:langevin}.

\subsection{The Rescaled Process}
We start by performing a diffusive rescaling to the equations of
motion (\ref{eqn:langevin}): $t \rightarrow t/\epsilon^2$, $x
\rightarrow \frac{x}{\epsilon}$. Using the fact that $\beta(c \,
t) = \sqrt{c} \beta(t)$ in law we obtain:
\begin{equation}
\tau \epsilon^2 \ddot{x} = \frac{1}{\epsilon}
v(\frac{x}{\epsilon}) - \dot{x} + \sigma \dot{\beta}.
\nonumber
\end{equation}
Introducing $y=\sqrt \tau \epsilon \dot{x}$ and $z=x/\epsilon$
we write this equation as a first order system:
\begin{equation}
\begin{array}{ccc}
\dot{x} & = &\frac{1}{\sqrt{\tau} \epsilon} y,\\
\dot{y} & = &\frac{1}{\sqrt{\tau} \epsilon^2} v(z) -
\frac{1}{\tau \epsilon^2} y + \frac{\sigma}{\sqrt{\tau} \epsilon}
\dot{\beta},\\
\dot{z} & = &\frac{1}{\sqrt{\tau} \epsilon^2} y
\end{array}
\label{eqn:langevin_rescaled}
\end{equation}
with the understanding that $z \in {\mathbb T}^d$ and $x,y \in {\mathbb R}^d.$ This
clearly exhibits the fact that the problem possesses two time scales of
${\cal O}(\epsilon)$ and ${\cal O}(\epsilon^2).$ We now
average out the fastest scales, on which $(y,z)$, evolve, and show that
the fast and large fluctuations in $x$ induces diffusion on time-scales
of ${\cal O}(1).$
\subsection{Multiscale Expansion}
The backward Kolmogorov equation associated to equations
(\ref{eqn:langevin_rescaled}) is
\begin{eqnarray}
\frac{\partial u^{\epsilon}}{\partial t} & = &
\frac{1}{\sqrt{\tau}\epsilon} y \cdot \nabla_x u^{\epsilon} +
\frac{1}{\epsilon^2} \left( \frac{1}{\sqrt{\tau}} y \cdot \nabla_z
+ \frac{1}{\sqrt{\tau}} v(z) \cdot \nabla_y + \frac{1}{\tau}
\mathcal{L}^{OU} \right) u^{\epsilon}.
    \nonumber \\ & := &
        \left( \frac{1}{\epsilon^2} \mathcal{L}_0 +
        \frac{1}{\epsilon} \mathcal{L}_1  \right)  u^{\epsilon},
\label{eq:backw_kolmog_2}
\end{eqnarray}
where:
\begin{align*}
\LL^{OU} & =  -  y \cdot \nabla_y + \frac{\sigma^2}{2} \Delta_y,\\
\mathcal{L}_0 & = \frac{1}{\sqrt{\tau}} \left( y \cdot \nabla_z +
v(z) \cdot \nabla_y \right) + \frac{1}{\tau} \LL^{OU},\\
\mathcal{L}_1 & = \frac{1}{\sqrt{\tau}} y \cdot \nabla_x.
\end{align*}
Note that $\LL^{OU}$ is the generator of a standard $d$--dimensional
Ornstein-Uhlenbeck process
 \cite[ch. 3]{Gar85}.
This process is ergodic with Gaussian invariant density
satisfying
\begin{equation}
(\cLo)^* \rou=0.
\label{eqn:rou}
\end{equation}
In order to carry out the analysis which follows we will make use
of the ergodic properties of the solution to \eqref{eqn:langevin}
with $x \to z$.
Using the tools developed in \cite{MatSt02} one can prove
that the process $z(t), \, y(t)$ with $y(t) = \sqrt{\tau}
\dot{z}(t)$ is ergodic on $\mathbb{T}^d \times \mathbb{R}^d$.\footnote{This
pair $z,y$ is the same as $z,y$ solving \eqref{eqn:langevin_rescaled}
up to a rescaling in time which is irrelevant to the ergodicity discussion here.}
The analysis implies that there exists a unique invariant density
$\rho(y,z)$ with support of positive measure on $\mathbb{T}^d \times
\mathbb{R}^d.$ The hypo-ellipticity of ${\cal L}_0^*$ established in
\cite{MatSt02} shows that the density is smooth and is hence the unique
solution to the stationary  Fokker--Planck equation associated to the process
\eqref{eqn:langevin}  \cite[ch. 11]{lasota}:
\begin{equation}
\LL_0^* \rho(y,z) :=  -\frac{1}{\sqrt{\tau}} \left( y \cdot
\nabla_z \rho + v(z) \cdot \nabla_y \rho \right) + \frac{1}{\tau} \left(
 \nabla_y \cdot (y \rho) + \frac{\sigma^2}{2} \Delta_y \right) \rho = 0.
\label{eqn:fp_sec}
\end{equation}
The Fokker--Planck operator $\LL_0^*$ is the adjoint of the
generator of the process $\LL_0$. The null space of the generator
$\LL_0$ consists of constants in $z, \, y$. Moreover, the
equation $\mathcal{L}_0 f = g$, has a unique (up to constants)
solution if and only if
\begin{equation}
\label{eq:int0}
\langle g \rangle_{\rho} := \int_{\R^d} \int_{\T^d}
g(y,z) \rho(y,z) \, dy dz = 0.
\end{equation}
In Appendix \ref{sec:app} we prove the ergodicity of the process $\{z, y \}$,
together with the fact that $\LL_0$ satisfies the Fredholm alternative.

We will assume that the average of the velocity with respect to
the invariant density $\rho$ vanishes:
\begin{equation}
 \langle v(z) \rangle_{\rho}=0.
\label{eqn:centering}
\end{equation}
From the identity $ \int_{\mathbb{T}^d} \int_{\mathbb{R}^d} y
\mathcal{L}_0^* \rho(y,z) \, dy dz = 0 $ and after an integration by
parts using \eqref{eqn:fp_sec},  it follows that condition
\eqref{eqn:centering} is equivalent to
\begin{equation}
 \langle y \rangle_{\rho} = 0.
\nonumber
\end{equation}
We assume that the following ansatz for the solution
$u^{\epsilon}$ holds:
\begin{equation}
u^{\epsilon} = u_0 + \epsilon u_1 + \epsilon^2 u_2 + \dots
\label{eq:expansion}
\end{equation}
with $u_i = u_i(x, y, z, t), \, i=1,2, \dots$. We substitute
(\ref{eq:expansion}) into (\ref{eq:backw_kolmog_2}) and obtain the
following sequence of equations:
\begin{equation}
\begin{array}{ccc}
\mathcal{L}_0 \, u_0 & = &0,\\
\mathcal{L}_0 \, u_1 & = &- \mathcal{L}_1 \, u_0,\\
\mathcal{L}_0 \, u_2 & = &- \mathcal{L}_1 \, u_1 +
\frac{\partial u_0}{\partial t}.
\end{array}
\label{eq:oep}
\end{equation}
From the first equation in (\ref{eq:oep}) we deduce that $u_0 =
u_0(x,t)$, since the null space of $\LL_0$ consists of functions which are
constants in $y$ and $z$. Now the second equation in (\ref{eq:oep}) becomes:
\begin{equation}
\mathcal{L}_0 u_1 = - \frac{1}{\sqrt{\tau}}y \cdot \nabla_x u_0.
\nonumber
\end{equation}
The centering condition that we have imposed on the vector field $v(y,z)$ implies
that $\langle y \rangle_{\rho} = 0$. Hence the above equation is well--posed. We
solve it using separation of variables:
\begin{equation}
u_1 = \Phi(y,z) \cdot \nabla_x u_0
\nonumber
\end{equation}
with
\begin{equation}
\mathcal{L}_0 \Phi(y,z) = - \frac{1}{\sqrt{\tau}}y.
\label{eqn:cell}
\end{equation}
This is the cell problem which is posed on $\T^d \times \R^d$. Now
we proceed with the third equation in \eqref{eq:oep}. We apply the solvability
condition to obtain:
\begin{eqnarray*}
\frac{\partial u_0}{\partial t}
                         &  =  &
                           \langle \mathcal{L}_1 u_1 \rangle_{\rho} \\
 & = & \frac{1}{\sqrt{\tau}} \langle
                             y_i \Phi_j \rangle_{\rho}
                             \pddd{u_0}{x_i}{x_j}\\
 & = & \frac{1}{\sqrt{\tau}} \langle
                             y \otimes \Phi \rangle_{\rho}:
                             D^2_x u_0.
\end{eqnarray*}
This is the Backward Kolmogorov equation which governs the dynamics on large
scales. We write it in the form
\begin{equation}
\frac{\partial u_0}{\partial t} =  \KK_{ij}
                             \pddd{u_0}{x_i}{x_j}
\label{eqn:homog_lang}
\end{equation}
where the effective diffusivity is
\begin{equation}
\mathcal{K}_{ij} = \frac{1}{\sqrt{\tau}} \langle  y_i \Phi_j
\rangle_{\rho}.
\label{eqn:eff_diff}
\end{equation}
The calculation of the effective diffusivity requires the solution
of the cell problem (\ref{eqn:cell}). Notice that the cell problem
is not elliptic -- it is, however, hypo-elliptic. This follows
from the calculations in Lemma \ref{lem:hypoell}.

Before studying various properties of the effective diffusivity, let us briefly present the
basic ingredients of the rigorous proof of the homogenization theorem presented in
\cite{per_hom_hypoell}. The basic idea is to apply
It\^{o} formula to the solution $\Phi(y,z)$ of the cell problem to obtain,
from \eqref{eqn:langevin_rescaled},
\begin{eqnarray}
x(t) & = & x(0) +  \frac{1}{\eps\sqrt{\tau}}\int_0^{t} y(s)
                        \, ds
          \nonumber \\ & = &
                x(0) - \eps \left( \Phi(y(t),z(t)) - \Phi(y(0),z(0))
         \right) + \frac{\sigma}{ \sqrt{\tau}} \int_0^{t}
         \nabla_y \Phi \left(y(s),z(s) \right) \, d \beta(s).
    \nonumber
\end{eqnarray}
It is straightforward to show that the bracketed terms on the right hand side of the
above equation converge to $0$, as $\eps \rightarrow 0$. By using the fact that the
$(y,z)$ process is fast, the martingale central limit theorem  facilitates proof
that the stochastic integral which appears in the above equation converges to a
Brownian motion whose covariance is given by the limit of the quadratic variation of
the stochastic integral \cite{kipnis}, \cite[Thm. 7.1.4]{ethier_86}. The application
of this theorem automatically provides us with the well--posedness of the limiting
Backward Kolmogorov equation. In the next section we establish this well--posedness
directly, through the multiple-scales framework.
%
%
\section{Properties of the Effective Diffusivity Tensor}
\label{sec:eff_diff_prop}
\subsection{Well-posedness of the Limiting Equation}
In the previous section we showed that the dynamics of the
inertial particles at large scales is governed by the backward
Kolmogorov equation \eqref{eqn:homog_lang}. In this  subsection we
prove that this equation is well--posed, i.e. that the effective
diffusivity is nonnegative. To see this, we first observe
that
\begin{equation}
\mathcal{L}_0^* (f \rho) = - \rho \mathcal{L}_0 f +
\frac{\sigma^2}{\tau} \rho \Delta_y f + \frac{\sigma^2}{\tau}
\nabla_y \rho \cdot \nabla_y f,
\nonumber
\end{equation}
for every periodic function $f(y,z)$ which is sufficiently smooth. We use
this equation, together with an integration by parts and some
algebra to obtain:
\begin{eqnarray}
\int_{\mathbb{R}^d} \int_{\mathbb{T}^d} f
           \left( \mathcal{L}_0 f  \right) \rho \, dy dz
              & = &
 - \frac{\sigma^2}{2 \tau} \int_{\mathbb{R}^d} \int_{\mathbb{T}^d}
|\nabla_y f |^2  \rho \, dy dz = - \frac{\sigma^2}{2 \tau} \langle
|\nabla_y  f|^2 \rangle_{\rho}.
\label{eqn:gen_negative}
\end{eqnarray}
This is the analogue of \eqref{eqn:gen_negative_first_ord} for passive tracers. Now
let $\phi = a \cdot \Phi$ where $\Phi $ is the solution of the cell problem
\eqref{eqn:cell} and $a$ is a constant vector in $\mathbb{R}^d$. The scalar quantity
$\phi$ satisfies the equation
\begin{equation}
\label{eqn:cell12}
\mathcal{L}_0 \phi = - \frac{1}{\sqrt{\tau}} a \cdot y.
\end{equation}
From the formula for the effective diffusivity, together with equation
(\ref{eqn:gen_negative}), we obtain:
\begin{eqnarray}
a . \mathcal{K}a & = &\frac{1}{\sqrt{\tau}}
\int_{\mathbb{R}^d}\int_{\mathbb{T}^d}
                        (a \cdot y)  (a \cdot \Phi)  \rho \, dy dz
         \nonumber  \\   & = &
                       -  \int_{\mathbb{R}^d} \int_{\mathbb{T}^d}
                        \phi \mathcal{L}_0 \phi  \rho \, dy dz
         \nonumber \\ & = &
                        \frac{\sigma^2 }{2 \tau} \int_{\mathbb{R}^d}
                       \int_{\mathbb{T}^d} |\nabla_y \phi |^2
                       \rho \, dy dz
         \nonumber \\ & = &
                        \frac{\sigma^2 }{2 \tau}
                        \langle |\nabla_y \phi|^2 \rangle_{{\rho}}
                        \geq 0.
\nonumber
\end{eqnarray}
%

%
%
%
%
\subsection{Alternative Representation of the Effective Diffusivity}
%
%
The aim of this subsection is the derivation of an alternative
representation for the effective diffusivity along the direction
of the vector $a$ in $\R^d$. To this end, we define the field
$\chi$ through $\phi = \sqrt{\tau} y . a + \chi$ with $\phi = \Phi
. a$, $\Phi$ being the solution of the cell problem
(\ref{eqn:cell}). Substituting this expression in \eqref{eqn:cell}
 we obtain the following modified cell problem:
\begin{equation}
\mathcal{L}_0 \chi = - a \cdot v.
\label{eqn:cell_2}
\end{equation}
The effective diffusivity along the direction of the vector $a$
expressed in terms of $\chi$ is:
\begin{equation}
a . \mathcal{K} a=  \frac{\sigma^2}{2} |a|^2 + \frac{\sigma^2}{
 2\tau} \langle  |\nabla_y \chi |^2 \rangle_{\rho} +
\frac{\sigma^2}{ \sqrt{\tau}} \langle \nabla_y \chi \cdot a\rangle_{\rho}
\label{eqn:eff_diff_2}
\end{equation}
Equations \eqref{eqn:cell_2} and \eqref{eqn:eff_diff_2} have
the same structure as the corresponding equations \eqref{eqn:cell1},
\eqref{eqn:thistwo} for the first order dynamics.
We will exploit this in the next section  when we consider
the $\tau \to 0$ limit of \eqref{eqn:eff_diff_2}.
%
%
%
%
\subsection{Incompressible Flows}
We are unable to prove the analogue of what is known for passive tracers,
namely that diffusion is always enhanced for incompressible flow fields.
Numerical evidence, however, suggests that this enhancement is seen
in a wide variety of situations and that, furthermore, the presence
of inertia further enhances the diffusivity over the passive tracer
enhancement.  See section \ref{sec:num_exp}.
%
%
%
%
\subsection{Potential Flows}
From the representation \eqref{eqn:eff_diff_2} we can prove that, for potential
flows, the diffusivity is  depleted for all $\tau > 0$, as is true for the
case $\tau=0$ of passive tracers. As for passive tracers we use the fact
that the explicit form of the invariant measure is known for potential
flows. From equation \eqref{eqn:gen_negative}, with $f = \phi$, the
facts that $\langle |a . y|^2 \rangle_{\rho} = \frac{\sigma^2}{2} |a|^2$ and
that $a. \nabla_y \rho = - \frac{2}{\sigma^2} \rho y . a$ and an
integration by parts we obtain:
\[
\frac{\sigma^2}{2 \sqrt{\tau}} \langle
 \nabla_y \chi \cdot a \rangle_{\rho} =a \cdot \KK a - \frac{\sigma^2}{2} |a|^2  
\]
We use the above formula in equation \eqref{eqn:eff_diff_2} to deduce
that, for potential flows, we have $\langle a . \nabla_y \chi \rangle_{\rho} =
-\frac{1}{\sqrt{\tau}} \langle |\nabla_y \chi|^2 \rangle_{\rho}$ which implies:
\begin{equation}
a \cdot \KK  a = \frac{\sigma^2}{2} |a|^2 - \frac{\sigma^2}{2 \tau} \langle
 |\nabla_y \chi |^2 \rangle_{\rho}.
\label{e:depl_grad}
\end{equation}
Hence, transport is always depleted for potential flows. This formula
should be compared with the formula \eqref{eq:square}
arising in the case $\tau=0$, passive tracers. We also remark that a more sophisticated
analysis, based on the variational formulation of the effective diffusivity for passive
tracers, yields that for potential flows, and at least in one dimension, the
diffusivity for $\tau >0$ is depleted even beyond its depletion for 
$\tau = 0$ \cite[Thm. 5.1]{olla_notes_94}:
$$
\KK(\tau) \leq \KK(\tau = 0) \leq \frac{\sigma^2}{2}.
$$
This is a sharper upper bound on $\KK(\tau)$ than the one that follows from  
\eqref{e:depl_grad}. On the other hand, equation \eqref{e:depl_grad} has the advantage
that it provides us with an explicit expression for the difference between molecular
diffusivity and $\KK(\tau)$ in terms of the solution of equation \eqref{eqn:cell_2}. 
%
%
%
%
%
%
\subsection{Conditions for the Centering Hypothesis
\eqref{eqn:centering} to be satisfied}
\label{subsec:centering}
In this section we present some conditions which ensure that the centering condition
\eqref{eqn:centering}. It is easy to check that this condition is satisfied for
potential flows. Indeed, the explicit form of the invariant density is known in this
case and this enables us to perform the following computation:
\begin{eqnarray}
\int_{\T^d} \int_{\R^d} v(z) \rho(y,z) \, dy dz
                         & = &
- \frac{1}{Z} \frac{1}{(\pi  \sigma^2)^{\frac{n}{2}}}  \int_{\T^d} \int_{\R^d}
 \nabla V(z)  e^{-\frac{2}{\sigma^2}\left(\frac{1}{2}y^2 + V(z)  \right)   } \,
dy dz \nonumber \\
                         & = &
\frac{\sigma^2}{2} \frac{1}{Z} \frac{1}{(\pi  \sigma^2)^{\frac{n}{2}}}
\int_{\T^d} \int_{\R^d}
\nabla  \left( e^{-\frac{2}{\sigma^2}\left(\frac{1}{2}y^2 + V(z)  \right) }
\right) \, dy dz
              \nonumber \\  & = & 0.
\nonumber
\end{eqnarray}
In the general case, for which the invariant density is not explicitly known, we have
to study the symmetry properties of the drift term in order to identify other classes
of flows for which the centering condition is satisfied. Let us consider the case of a
parity invariant flow, i.e. a flow satisfying the condition
\begin{equation}
v(-z) = - v(z).
\label{eqn:parity}
\end{equation}
It follows from \eqref{eqn:parity} that the solution of equation
\eqref{eqn:fp_sec} --i.e. the invariant density-- satisfies
$$
\rho(y,z) = \rho(-y, -z).
$$
Hence \eqref{eqn:centering} is satisfied.

As a concrete example which shows that condition \eqref{eqn:parity} is necessary for
the centering hypothesis \eqref{eqn:centering} to be satisfied, let us consider the
two--dimensional velocity field
\begin{equation}
v(z_1,z_2) = (\sin z_1, - \cos z_1).
\label{e:pot_vel}
\end{equation}
We have that
$$v_1(-z_1) = - v_1(z_1) \quad \mbox{and} \quad v_2(-z_1) = v_2(z_1),$$
\begin{figure}
\begin{center}
\includegraphics[width=2.9in, height = 2.9in]{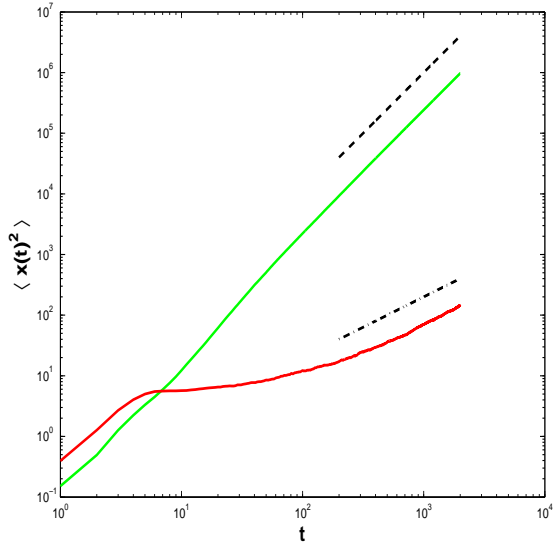}
\caption{ Second moments of the particle position versus time for the velocity field
\eqref{e:pot_vel}. The graphs of $t$ and $t^2$ are also included for comparison.}
\end{center}
\label{fig:pot}
\end{figure}
and hence we expect diffusive long time behavior along the $z_1$ direction and
ballistic motion along the $z_2$ direction. In Figure 3 we present the second
moments $\langle x_1(t)^2 \rangle, \, \langle x_2(t)^2\rangle$ of solutions to
equation \eqref{eqn:langevin} with $v(x)$ given by \eqref{e:pot_vel}\footnote{The
moments are obtained through Monte Carlo simulations. The details of the numerical
simulations presented in this paper are discussed in section \ref{sec:num_exp}.}.
%
\subsection{Homogenization When the Centering Hypothesis
\eqref{eqn:centering} is not Satisfied}
\label{subsec:ballistic}
In the previous section we imposed the centering condition \eqref{eqn:centering} in
order to ensure that there is no mean drift and that the motion of the inertial
particles at long scales is diffusive. In the case where this condition is not
satisfied, then we expect that the effective behavior of the particles is described
by a transport equation and that the diffusion appears only as a higher order
correction. Indeed, an analysis similar to the one presented in the previous
section, using the advective rescaling $t=t/\epsilon$ and $x=x/\epsilon,$ shows that
in this case the particle motion at large scales is goverened by the following
backward Kolmogorov equation
\[
\pd{u}{t} = \langle  v_i \rangle_{\rho} \pd{u}{x_i} +
\epsilon \KK_{ij} \pddd{u}{x_i} {x_j},
\]
where $\langle v \rangle_{\rho} = \int_{\T^d} \int_{\R^d} v(z) \rho(y,z) \, dy
dz$. Alternatively, the behavior of the particles is diffusive at the reference
frame moving with the mean flow.
%
%
%
%
\subsection{The Homogenization Problem for Equation \eqref{eqn:crisanti}}
%
%
%
Let us now consider the equation \eqref{eqn:crisanti} in the
presence of molecular diffusion. For steady flows this equation
becomes
\begin{equation}
 \ddot{x} = \delta v(x) \cdot \nabla v(x) + \frac{1}{\tau} \left( v(x) -
            \dot{x} + \sigma \dot{\beta} \right),
\label{eqn:maxey_noise}
\end{equation}
where $\delta$ is the ratio of fluid density to particle density. The techniques
developed in \cite{MatSt02} enable us to conclude that there exists a unique,
smooth, invariant density for the process $\{x(t), \, \sqrt{\tau} \dot{x}(t) \}$,
which we denote by $\rho^{\delta}(y,z)$. It is straightforward to check that the
assumption $v(-z) = - v(z)$ ensures that the drift term $b^{\delta}(z) = \delta v(z)
\cdot \nabla v(z) + \frac{1}{\tau} v(z)$ is centered with respect to this invariant
density, $\langle  b^{\delta} \rangle_{\rho^{\delta}}=  0$. Hence, the multiscale
techniques developed in section \ref{sec:homog} apply with $\frac{1}{\tau}v(z) \to
b^{\delta}(z)$ and we can conclude that  the rescaled process $x^{\epsilon} =
\epsilon x(t/\epsilon^2)$, with $x(t)$ being the solution of equation
\eqref{eqn:maxey_noise}, converges as $\epsilon \rightarrow 0$ to a Brownian motion.
The covariance--effective diffusivity-- of this Brownian motion is
\begin{equation}
\KK^{\delta} = \frac{1}{\sqrt{\tau}} \langle \Phi^{\delta} \otimes y
               \rangle_{\rho^{\delta}}
\nonumber
\end{equation}
where $\Phi^{\delta}$ solves the cell problem
\begin{equation}
\LL^{\delta} \Phi^{\delta} = -\frac{1}{\sqrt{\tau}} y,
\label{e:cell_delta}
\end{equation}
with
\[
\LL^{\delta} = \left(\sqrt{\tau} b^{\delta} \cdot \nabla_y + \frac{1}
               {\sqrt{\tau}} y \cdot \nabla_x \right) + \frac{1}{\tau} \LL_{OU}.
\]
Now the effective diffusivity depends on $\delta$ as well as on $\tau$ and $\sigma.$
We study the effect of varying $\delta$ in section \ref{sec:num_exp}, by means of
numerical experiments. The main interest here stems from the fact that, for
$\sigma=0$, equation \eqref{eqn:crisanti} exhibits effective diffusive behaviour for
certain cellular flows and choices of $\delta.$  How this diffusive mechanism
interacts with molecular diffusion is a matter of some interest.

\section{Small $\tau$ Asymptotics}
\label{sec:stau}

In this section we show that the effective diffusivity tensor $\KK=\KK(\tau)$
reduces to the effective diffusivity tensor from first order dynamics, as $\tau \to
0$. To this end we define
$$\cA_0=\cLo,\quad \cA_1=y\cdot \nabla_z+v(z)\cdot \nabla_y$$
and then, by polarization,
the effective diffusion tensor $\KK(\tau)$ is determined by
\eqref{eqn:eff_diff_2} with $\chi$ solving \eqref{eqn:cell_2},
$\rho$ solving \eqref{eqn:fp_sec} and
$$\cL_0=\frac{1}{\tau}\cA_0+\frac{1}{\sqrt \tau}\cA_1.$$
We wish to expand $\chi$ and $\rho$ in powers of $\tau$ and show that the leading
order behaviour of $\KK(\tau)$ is given by \eqref{eqn:thistwo}. Higher order terms
in the small $\tau$ expansion of the effective diffusivity $\KK(\tau)$ will be
computed only for the case of a one dimensional potential flow in section
\ref{subsec:1d_pot}. Perturbation calculations similar to the ones presented in this
section were reported in \cite{SubBr04_2, SubBr04_1} for linear shear flows.

\subsection{Expansion for $\chi$}
We set
$$\chi=\chi_0+\sqrt \tau \chi_1+\tau \chi_2 +\cdots$$
in \eqref{eqn:cell_2} and then find
\begin{align}
\cA_0 \chi_0 &=0\\
\cA_0 \chi_1 &=-\cA_1 \chi_0\\
\cA_0 \chi_2 &= -a\cdot v-\cA_1\chi_1.
\end{align}
The first of these equations implies that $\chi_0=\chi_0(z)$ only and
the second is soluble because
$$\langle \cA_1 \chi_0 \rangle_{\rou}=\langle y\cdot \nabla_z
\chi_0(z) \rangle_{\rou}=0.$$ Here $\rou$ is the mean zero Gaussian invariant
density of the OD process, satisfying \eqref{eqn:rou}. Solving for $\chi_1$ gives
$$\chi_1=y \cdot \nabla_z \chi_0(z) + \psi_1(z)$$
and the solvability condition for the $\chi_2$ equation yields
$$\frac{\sigma^2}{2}\Delta_z \chi_0+v(z)\cdot \nabla_z \chi_0=-a.v(z).$$
Thus  $\chi_0$ is  the solution of the cell problem arising in the passive tracer
case, equation \eqref{eqn:cell1}. Furthermore $\nabla_y \chi_1=\nabla_z \chi_0(z)$
and hence
\begin{equation}
\label{eqn:A}
\nabla_y \chi=\sqrt \tau \nabla_z \chi_0(z)+{\cal O}(\tau).
\end{equation}
Notice that the function $\psi_1(z)$ is undetermined at this point, but
that it does not enter equation \eqref{eqn:A}.
\subsection{Expansion For $\rho$}
Now it remains to expand $\rho(y,z)$ from \eqref{eqn:fp_sec} in $\tau.$
Notice that
$$\cA_{0}^* \rou(y)=0,\quad \nabla_y \rou(y)=-2y\rou(y)/\sigma^2.$$
Then, by \eqref{eqn:what} with $z \to y$ and $\cL_0 \to \cA_0$
we find that
\begin{align*}
\cA_0^*(f\rou) & = -\rou[\cA_0f-\sigma^2 \Delta_y f+2y\cdot \nabla_y f]\\
& = \rou \cA_0 f.
\end{align*}
Let $\rho=\rou \pit.$ Then
\begin{align*}
0&=\cL_0^*(\rou \pit)\\
&=\frac{1}{\tau}\cA_0^*(\rou \pit)+\frac{1}{\sqrt \tau}\cA_1^*(\rou \pit)\\
&=\frac{\rou}{\tau}\cA_0\pit+\frac{\rou}{\sqrt \tau}
\left( y\cdot \nabla_z \pit+v\cdot\nabla_y \pit-\frac{2}{\sigma^2}(v\cdot y)\pit\right).
\end{align*}

Since $\rou>0$ everywhere we may divide through by it in the above
expression. If we then set
$$\pit=\pi_0+\sqrt \tau \pi_1+\tau \pi_2$$
we find that
\begin{align*}
-\cA_0 \pi_0 &=0\\
-\cA_0 \pi_1 &=
y\cdot \nabla_z \pi_0+v\cdot\nabla_y \pi_0-
\frac{2}{\sigma^2}(v\cdot y)\pi_0\\
-\cA_0 \pi_2 &=
y\cdot \nabla_z \pi_1+v\cdot\nabla_y \pi_1-
\frac{2}{\sigma^2}(v\cdot y)\pi_1.
\end{align*}

The first equation shows that $\pi_0=\pi_0(z)$ only and the second
then gives
$$\pi_1(y,z)=y\cdot\{\nabla_z \pi_0-\frac{2}{\sigma^2}\pi_0 v\} - \mu_1.$$
As with the second term in the dual expansion, the function $\mu_1$ is undetermined at
this point. However, as the subsequent calculations will show, it is not needed in order to
compute the first term in the small $\tau$ expansion of the effective diffusivity.

The negative of the right hand side of the third equation is
\begin{eqnarray*}
&&-y \otimes y:\{D_z^2\pi_0-\frac{2}{\sigma^2}\nabla_z\{\pi_0 v\}\}
-v\cdot \{ \nabla_z \pi_0-\frac{2}{\sigma^2}\pi_0 v\}
\\ &&+y \otimes y:\{\nabla_z \pi_0-\frac{2}{\sigma^2}\pi_0 v\}\otimes\frac{2v}
{\sigma^2}  + y \cdot \{ \nabla_z \mu_1-\frac{2}{\sigma^2} v \mu_1 \}.
\end{eqnarray*}
Now
$$\langle y \otimes y \rangle_{\rho^{OU}}=\frac{\sigma^2}{2}I \quad \mbox{and} \quad
\langle y \rangle_{\rho^{OU}} = 0$$
and so the solvability condition for $\pi_2$ yields
$$-\frac{\sigma^2}{2}\Delta_z \pi_0+\nabla_z \cdot \{\pi_0 v\}=0$$
and hence shows that $\pi_0$ coincides with the passive tracer case,
equation \eqref{eqn:id1}.  In summary, we have shown that
\begin{equation}
\label{eqn:B}
\rho(y,z)=\rou(y)\piz(z)+{\cal O}(\sqrt \tau)
\end{equation}
where $\piz(z)$ is the invariant density from the first order
dynamics, satisfying \eqref{eqn:id1}.

\subsection{Limit of the Diffusivity Tensor}

Combining \eqref{eqn:A} and \eqref{eqn:B} in \eqref{eqn:eff_diff_2}
gives
$$a \cdot\KK a = \frac{\sigma^2}{2}|a|^2+\frac{\sigma^2}{2}\int_{\mathbb{T}^d}
\pi(z)|\nabla_z \chi_0(z)|^2dz
+\sigma^2\int_{\mathbb{T}^d}\pi(z) a.\nabla_z \chi_0(z)+{\cal O}(\sqrt \tau)$$
which, to leading order in $\tau$,
is the expression for the first order dynamics.
\subsection{One dimensional Potential Flows}
\label{subsec:1d_pot}
The calculation of higher order terms in the small $\tau$ expansion for the
effective diffusivity is quite involved. Moreover, in the general case, these higher
order terms do not seem to be of definite sign. However, it is possible to compute
explicitly the next term in the small $\tau$ expansion and to prove that it has a
definite sign in the case of one dimensional potential flows and we now pursue this.
Consider the equation
\begin{equation}
\tau \ddot{z} = - V'(z) - \dot{z} + \sigma \dot{\beta}.
\label{e:1dgrad}
\end{equation}
In this case we only need to solve perturbatively the cell problem \eqref{eqn:cell} since
the
stationary Fokker--Planck equation corresponding to \eqref{e:1dgrad} is exactly solvable
yielding an invariant density independent of $\tau$:
\begin{equation}
\rho(y,z)  = \frac{1}{Z} \frac{1}{\sqrt{\pi \sigma^2}}
e^{-\frac{2}{\sigma^2}\left(\frac{1}{2}y^2 + V(z)  \right)   } ,
\label{eqn:gibbs}
\end{equation}
with $Z = \int_0^1 \exp(-\frac{2}{\sigma^2} V(z)) \, dz $ and $y =
\frac{1}{\sqrt{\tau}} \dot{z}$. The effective diffusivity--which
now is a scalar-- is given by the formula:
\begin{equation}
\mathcal{K} = \frac{1}{\sqrt{\tau}}  \int_{-\infty}^{+ \infty}
\int_0^1 y \phi(y,z) \rho(y,z) \, dy dz.
\label{eqn:eff_diff_grad}
\end{equation}
A lengthy calculation enables us to compute the first four terms in the small $\tau$
expansion for the cell problem:
\begin{subequations}
\begin{equation}
\phi_0 = \chi(z),
\end{equation}
\begin{equation}
\phi_1(y,z) = y (\chi'(z) +1),
\end{equation}
\begin{equation}
\phi_2(y,z) = \frac{1}{2} y^2 \chi''(z) + \psi_2(z),
\end{equation}
\begin{equation}
\phi_3(y,z) =  \frac{1}{6} y^3  \chi'''(z) + y \left(\frac{1}{2} \sigma^2 \chi'''(z) -
V'(z) \chi''(z) + \psi_2'(z) \right).
\end{equation}
\label{e:exp_grad}
\end{subequations}
The cell problem for the first order dynamics, equation \eqref{eqn:cell_1d}, can be
solved explicitly \cite{vergassola} to give
$$
\chi(z) = \int_0^z \widehat{\rho}(z) \, dz - z + c_0,
$$
with
$$
\widehat{\rho}(z) = e^{\frac{2}{\sigma^2}V(z)} , \quad
\widehat{Z} = \int_0^1 e^{\frac{2}{\sigma^2}V(z)} \, dz.
$$
We use the above formula for $\chi(z)$ in equation \eqref{eqn:eff_diff_1d} to obtain the
following formula for the effective diffusivity
\begin{equation}
\label{eqn:ptf} \mathcal{K}(\tau = 0) = \frac{\sigma^2}{2} \frac{1}{Z \widehat{Z}}.
\end{equation}
The term $\psi_2(z)$ satisfies an equation similar to the cell problem of the first
order dynamics. The solution of this equation is
$$
\psi_2(z) = -\frac{2}{\sigma^2} \int_0^z \left(\frac{1}{2} (V'(z))^2 + \frac{3 \sigma^2}{4}
V''(z)  \right) \widehat{\rho}(z) \, dz + c_1 \int_0^z \widehat{\rho}(z) \, dz + c_2,
$$
where
$$
c_1 =  -\frac{2}{\sigma^2} \int_0^z \frac{1}{2} (V'(z))^2  \widehat{\rho}(z) \, dz.
$$
The values of the constants $c_0, \, c_2$ are not needed for the computation of the
effective diffusivity. We substitute now \eqref{e:exp_grad} and \eqref{eqn:gibbs},
using the formulas for $\chi(z)$ and $\psi_2(z)$ and for the moments of the
Ornstein--Uhlenbeck process. The final result is
$$
\mathcal{K} = \frac{\sigma^2}{2} \frac{1}{Z \widehat{Z}} - \tau \int_0^1
\frac{1}{2} (V'(z))^2  \widehat{\rho}(z) \, dz + {\mathcal O}(\tau^2).
$$
The above formula shows that, for $\tau$ sufficiently small, the diffusivity is depleted
beyond the depletion exhibited by homogenization in the passive tracer case,
which is given by \eqref{eqn:ptf}.

%
%
%
%
\section{Numerical Experiments}
\label{sec:num_exp}
In this section we study the dependence of the effective diffusivity for equations
\eqref{eqn:langevin}, \eqref{eq:velf} on the non--dimensional parameters of the
problem $\tau, \, \sigma$ and $\delta$. For simplicity all the experiments we
perform are for the Taylor--Green flow \eqref{e:tg}\footnote{In our derivation of
the homogenized equation and the formula for the effective diffusivity we assumed
that the velocity field is $1$--periodic, rather than $2 \pi$--periodic.
Of course the analysis, as well as the formulas that we derived, are trivially
extended to encompass this change of period.}
$$
 U(x) = \nabla^\bot \psi_{TG}(x), \quad \psi_{TG}(x) =  \sin(x_1)
\sin(x_2).
$$
The closed streamlines of Lagrangian particle paths in this velocity field is a
rather special situation and we describe numerical experiments for other stream
functions, including open streamline topologies, in \cite{BPS04}.

It is straightforward to check that the Taylor--Green flow satisfies condition
\eqref{eqn:parity} and hence the absence of ballistic motion at long scales is
ensured. Moreover, the symmetry properties of \eqref{e:tg} imply that the two
diagonal components of the effective diffusivity are equal, whereas the
off--diagonal components vanish. In the figures presented below we use the notation
$K:= \KK_{11}=\KK_{22}$.

Rather than solving the cell problem \eqref{e:cell_delta}, we compute the
effective diffusivity using Monte Carlo simulations: we solve the equations of
motion \eqref{eqn:maxey_noise} numerically for different realizations of the noise
and we compute the effective diffusivity through the formula
\begin{equation}
\mathcal{K} = \lim_{ t \rightarrow \infty} \frac{1}{2 t} \langle (x(t) - \langle
x(t) \rangle) \otimes (x(t) - \langle x(t) \rangle) \rangle,
\nonumber
\end{equation}
where $\langle \cdot \rangle$ denotes ensemble average. We solve the stochastic equations
of motion using Milstein's method, appropriately modified for the second order SDE
\cite[p. 386]{kloeden92}:
$$
x_{n+2} = (2- r) x_{n+1} - (1 - r) x_n + r \Delta t v(x_{n+1}) + \sigma r \Delta t
\mathcal{N}(0,1).
$$
where $r = \frac{\Delta t}{\tau}$ and $ s_1 = 1 - \frac{r}{2}, \;\; s_2 = 1 +
\frac{r}{2}. $. This method has strong order of convergence $1.0$.
\footnote{All experiments
reported in this paper have been independently verified by use of
an alternative, linearly implicit, method. The agreement between the
statistics computed using these two methods is excellent.}
We use $N = 1024$ uniformly distributed particles
in $2 \pi \mathbb{T}^2$ with zero initial velocities and we integrate over a very
long time interval (which is chosen to depend upon the parameters of the problem)
with $\Delta t \approx 5.10^{-4} \min \{1, \tau \}$.

In some instances we compare the effective diffusivities for inertial
particles with those for passive tracers. The latter are computed by solving
the cell problem directly, by means of a spectral method similar to that
described in \cite{mclaugh}, together with extrapolation into
parameter regimes where the dependence of the diffusivity is provably linear.

\begin{figure}
\begin{center}
\includegraphics[width=2.9in, height = 2.9in]{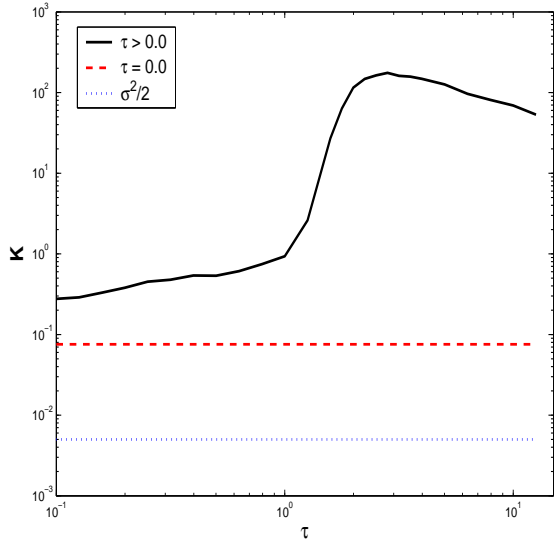}
\end{center}
\begin{center}
\caption{ Effective diffusivity versus $\tau$ for $\delta = 0.0, \sigma = 0.1$  .}
\label{fig:tau_k}
\end{center}
\end{figure}
%
%
%
%
\begin{figure}
\centerline{
\begin{tabular}{c@{\hspace{3pc}}c}
\includegraphics[width=2.9in, height = 2.9in]{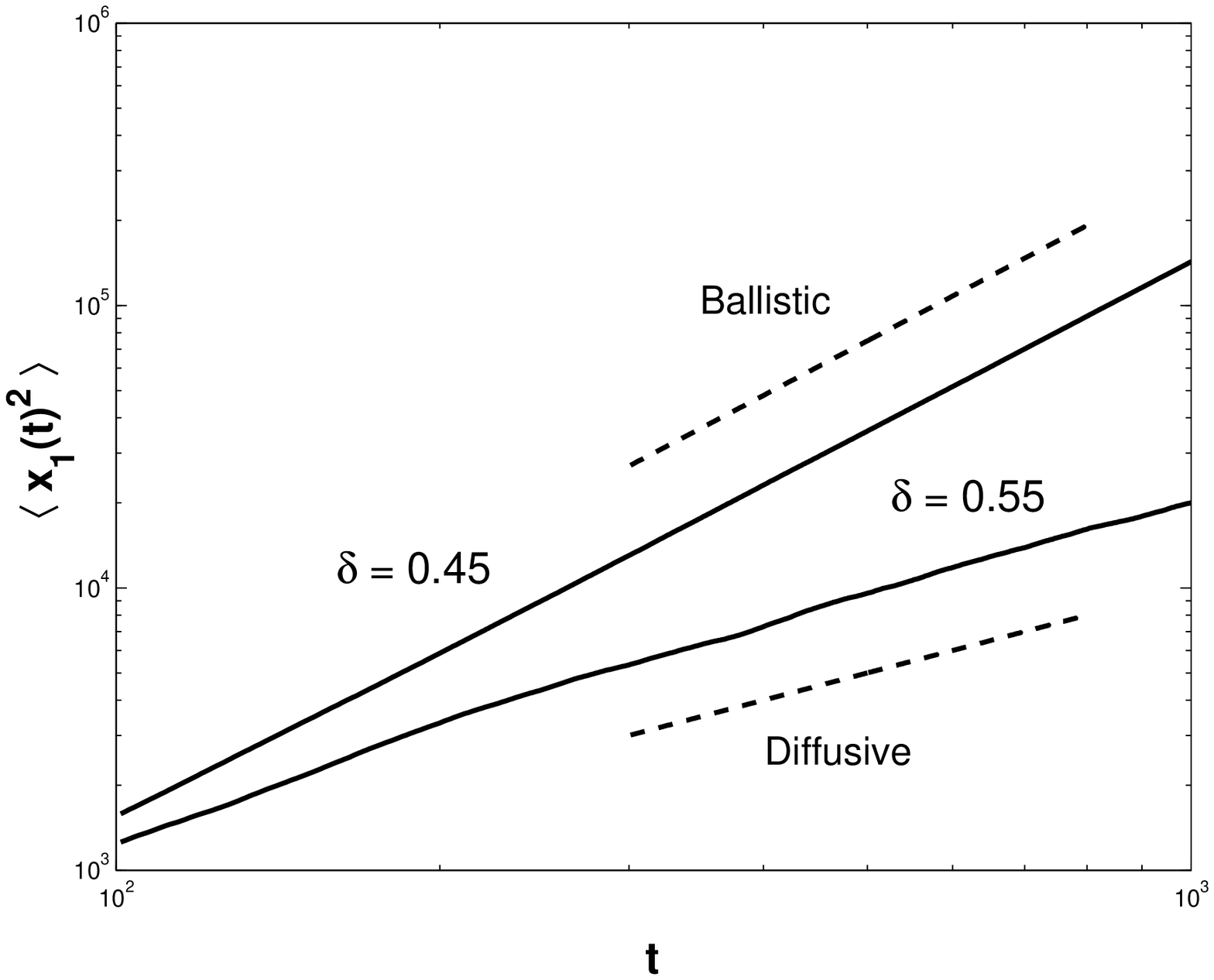} &
\includegraphics[width=2.9in, height = 2.9in]{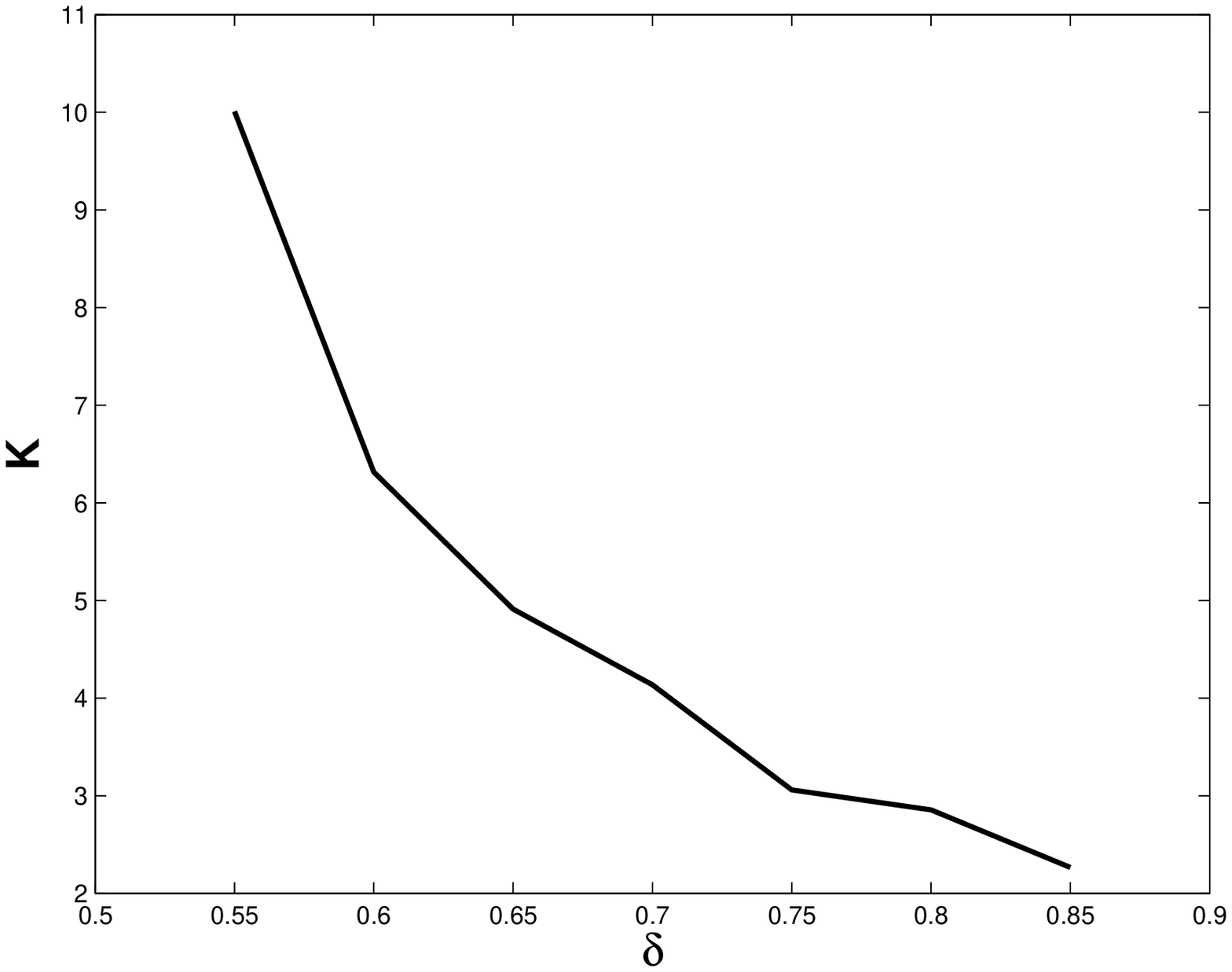} \\
a.~~  The second moment of the particle velocity for \\ $\delta =0.45$ and $\delta
=0.55$. \\ The lines $t$ and $t^2$ are also plotted for comparison. & b.~~ $\KK$ vs
$\delta$
\end{tabular}}
\begin{center}
\caption{  Effective diffusivity as a function of $\delta$ for $\sigma = 0.0$ and
$\tau = 2.0$.}
\label{fig:del2_k}
\end{center}
\end{figure}
\begin{figure}
\begin{center}
\includegraphics[width=2.9in, height = 2.9in]{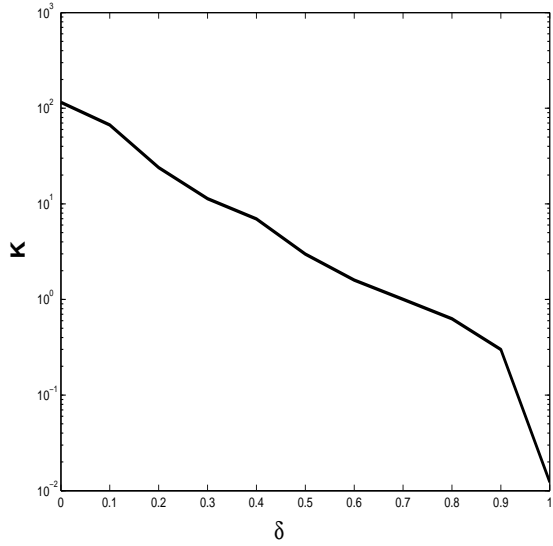}
\end{center}
\begin{center}
\caption{  Effective diffusivity as a function of $\delta$ for $\sigma = 0.1$ and
$\tau = 2.0$.}
\label{fig:del1_k}
\end{center}
\end{figure}
%
%
%
%
\begin{figure}
\centerline{
\begin{tabular}{c@{\hspace{3pc}}c}
\includegraphics[width=2.9in, height = 2.9in]{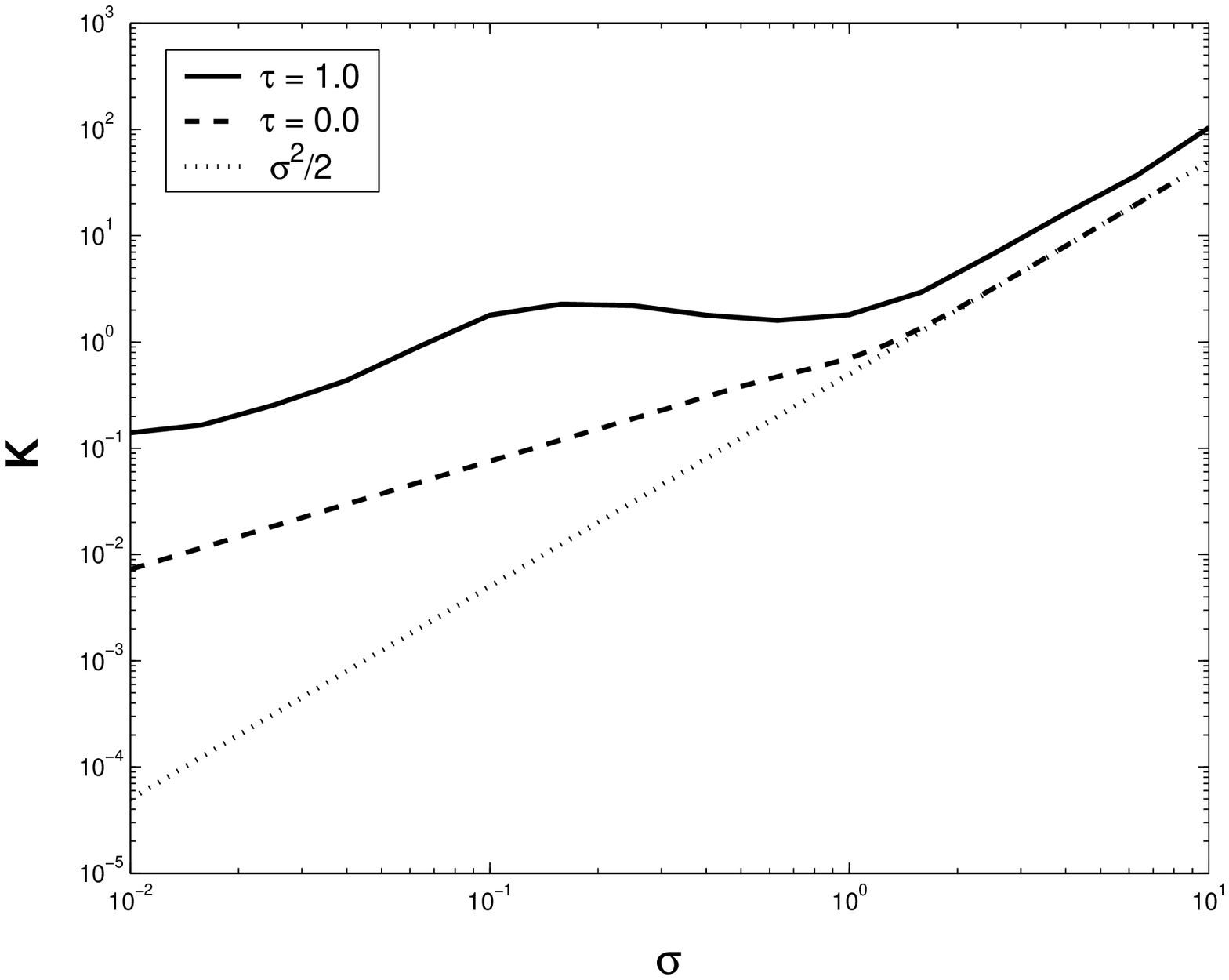} &
\includegraphics[width=2.9in, height = 2.9in]{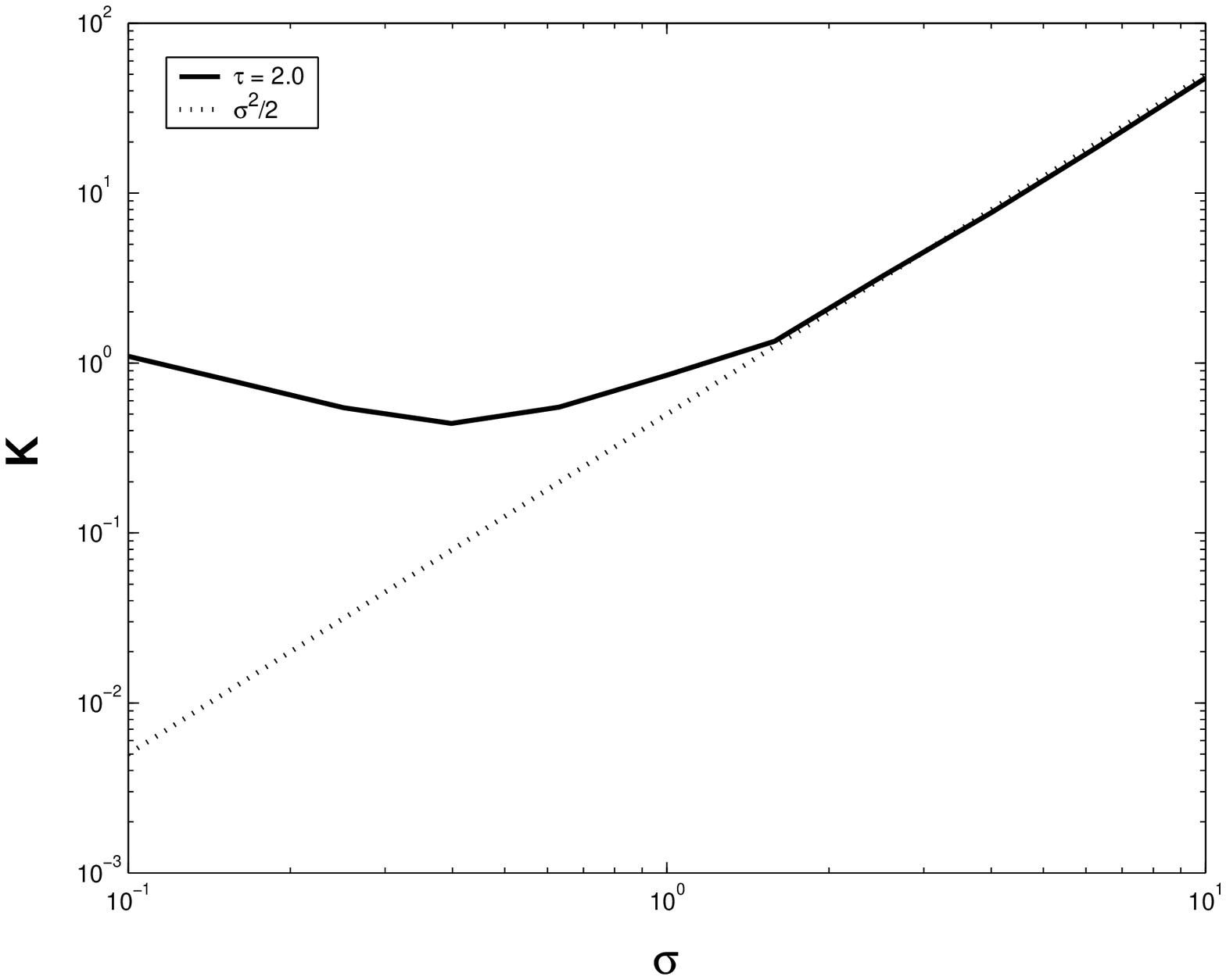} \\
a.~~  $\delta = 0.0, \, \tau = 1.0$  & b.~~ $\delta = 0.7, \, \tau = 2.0$
\end{tabular}}
\begin{center}
\caption{ Effective Diffusivity versus $\sigma$ for $\delta = 0.0$ and $\delta =0.7$
.}
\label{fig:sig_k}
  \end{center}
\end{figure}
\subsection{The Effect of $\tau$ on Diffusivity}
We compute the effective diffusivity as a function of the non--dimensional particle
relaxation  time $\tau$ for the Taylor--Green flow when $\delta =0.0$. Our results
are presented in Figure \ref{fig:tau_k}, for $\sigma = 0.1$. For comparison, the
effective diffusivity of the tracer particle ($\tau=0$)
and that of the free particle, namely
$\frac{\sigma^2}{2}$, are also plotted. The parameter $\tau$, apart from influencing
the effective diffusivity, introduces an additional time scale into the problem \cite{sancho_al04b}.
 In particular, for $\tau$ large, we need to integrate the equations of motion over a
longer time interval in order to compute accurate statistics.

The main interest in this data is that it shows highly non--trivial
dependence of the effective diffusivity on the parameter $\tau$
as well as giving quantative information about how inertia enhances
the diffusivity over that obtained in the case of passive tracers. We remark,
in particular, that the effective diffusivity reaches its maximum for 
$\tau = \OO(1)$. It is in this regime that the free-flight time-scale for the
inertial particle is of the same order as that induced by the velocity field.
A similar phenomenon, in a different context, has been observed by Vassilicos et. al. 
\cite{VassChenGotOsb_05}.

\subsection{The Effect of $\delta$ on Diffusivity}
It is a well documented result \cite{crisanti_inert, maxey_5, wang, maxey_3,
maxey_4} that the particle trajectories of \eqref{eqn:crisanti} perform an effective
Brownian motion even in the absence of noise, in certain parameter regimes. The
linear stability analysis of \eqref{eqn:crisanti} for the Taylor--Green flow
indicates that in the parameter regime $\delta \in (1/\tau, 1)$ we expect a very
complicated, chaotic behavior which might be interpreted as an effective Brownian
motion at long times. When $U(x) = \nabla^\bot \psi_{TG}(x)$ then the noise free
dynamics arising from \eqref{eqn:langevin} when $\sigma=0$ has two sets of
equilibria, $X_0^1 = [n \pi, 0, m \pi, 0], \, n,m \in \mathbb{N}$ and $X_0^2 =
[\frac{\pi}{2} + n \pi, 0, \frac{\pi}{2} + m \pi,0], \, n,m \in \mathbb{N}$. The
first set of equilibria have two dimensional unstable manifold for
$\delta>\tau^{-1}$ and one dimensional for $\delta<\tau^{-1}$; the second set have a
two dimensional unstable manifold for $\delta<1$ and is stable for $\delta>1.$
Numerically we observe diffusive behaviour, in the absence of noise, if and only if
$\delta \in (\tau^{-1},1).$ In this regime, all equilibria have two dimensional
unstable manifolds. The basic mechanism for diffusion is chaotic mixing
caused by separation near the separatrices of equilibria; it is unsurprising
therefore that the linear stability of equilibria play a strong role on determining
the interval of $\delta$ in which diffusion occurs.

In Figure \ref{fig:del2_k}a we plot the second moment of the particle position as a
function of time for values of $\delta$ below and above the threshold $\delta =
0.5$, and in the absence of noise. As expected, the particle motion is ballistic for
$\delta = 0.45$ and diffusive for $\delta = 0.55$.

In Figure \ref{fig:del2_k}b we plot the effective diffusivity as a function of
$\delta$, again in the absence of noise. Since we have chosen $\tau =2.0$ we expect an
effective Brownian motion for $\delta \in (0.5, 1)$. Notice that the effective diffusivity 
increases as $\delta \rightarrow 0.5^+$, and appears to diverge in the limit.
This is to be expected, since $\delta =0.5$ separates ballistic motion
(for $\delta < 0.5$) and diffusive motion (for $\delta > 0.5$) motion. 

In Figure \ref{fig:del1_k} we plot the effective diffusivity as a function of $\delta$ when
the particles are subject to additional molecular diffusion. In this case the effective diffusivity
is a decreasing function of $\delta$.

\subsection{The Effect of $\sigma$ on Diffusivity}
It is well known, see e.g. \cite{kramer}, that for the case of passive tracers the
effective diffusivity depends on the molecular diffusivity $\sigma$ in a highly non--linear,
 very complicated way. In particular, the limit as $\sigma$ tends to $0$ is singular and
the enhancement in the diffusivity--for divergence free flows-- depends crucially on
the topology of the streamlines. It is therefore interesting to study the dependence
of the effective diffusivity on $\sigma$ for the inertial particles problem.

In Figure \ref{fig:sig_k} we present the effective diffusivity as a function of the
molecular diffusion for two sets of parameters: (a) for $\delta =0.0$ (for
which the streamlines are closed), $\tau = 1.0$ and (b) for $\delta =0.7, \, 
\tau =2.0$ (a regime in which there exists a well defined effective diffusivity even in the
respectively. The diffusivity of the free particle $\frac{\sigma^2}{2}$ is
also plotted for comparison. Moreover, in Figure \ref{fig:sig_k}(a),
we also plot the effective diffusivity for passive tracers.

In both figures we see the clear enhancement of diffusivity over the bare
molecular value. Furthermore, in Figure \ref{fig:sig_k}(a) we also
observe that the enhancement in
 the diffusivity is significantly greater for $\tau >0$ (inertial particles)
than it is for $\tau = 0$ (passive tracers), especially when $\sigma$ is small.

A further interesting observation concerns the dependence of the diffusivity on
$\sigma$. For $\delta=0$ the dependence is highly non-trivial, exhibiting both a
local maximum and a local minimum. For $\delta=0.7$ the presence of inertia leads to
an effective diffusivity which increases as $\sigma$ becomes smaller.  This should
be contrasted with the established fact that, for passive tracers in periodic
cellular flows, the effective diffusivity decreases linearly in $\sigma$, for
$\sigma$ sufficiently small \cite{shraiman, solomon1}.  It would be interesting to
understand how the enhancement scales with $\sigma \rightarrow 0$ and to compare it
with known theoretical results for the passive tracer case; in particular, it would
be interesting to extend the theory of maximally and minimally enhanced diffusion
\cite{mclaugh} to the inertial situation studied here.

\section{Conclusions}
\label{sec:conclusions}
The problem of periodic homogenization for inertial particles is considered in
this paper. It is shown that, at long times and large scales, the inertial
particles perform Brownian motion and a formula for the effective diffusivity
is derived.  Furthermore, the dependence of the effective diffusivity on the
non--dimensional particle mass $\tau$ a(Stokes number)
and the ratio of the fluid to particle
density $\delta$ is studied, by means of analysis and numerics.

It is shown that, as $\tau \rightarrow 0,$ the effective diffusivity
converges to the one obtained from the homogenization of passive tracers.
Moreover, it is shown through numerical experiments that for a variety of
interesting divergence free flows, the diffusivity in the presence of inertia
is enhanced much beyond the well documented enhancement of the diffusivity for
passive tracers.  Furthermore, the dependence of the effective
diffusivity on $\tau$ and $\delta$ is studied numerically in some detail.

The calculation of the effective diffusion tensor
requires the numerical solution of equations \eqref{eqn:cell} and \eqref{eqn:fp_sec}. It is 
coneivable that this task might be as computationally demanding as direct Monte Carlo 
simulations, because the domain of the PDE is unbounded in the momentum
variable, and because the PDE is not elliptic (only hypo-elliptic).
This is to be contrasted to the case of passive tracers in 
periodic flows; there the calculation of the effective diffusivity requires 
the solution of the elliptic PDE \eqref{eqn:cell_1d} on a periodic domain.
Equations of this type can be routinely and effeciently solved using, 
for example, a spectral method. From this point of view our results might not
provide any computational advantage over Monte Carlo simulations. However,
the results reported in this paper provide a mathematical framework for 
rigorous analysis of the dependence of the effective diffusion coefficient
on the physical parameters of the problem. We have already undertaken such
an analysis to study the limit of small Stokes number and we plan to undertake
further studies in future work.

The numerical results reported in section \ref{sec:num_exp}, 
for a simple two-- dimensional steady flow, exhibit a wide range of interesting physical phenomena. 
As examples we mention the dependence of the effective diffusivity on the Stokes number and the 
fluid/particle density ratio, for a given streamline topology, question which are, we believe,
 of great interest to the applied community. The purpose of this work is
to develop a 
 framework within which such questions can be addressed. We plan to investigate
some of these issues in future work. The dependence of the
effective diffusivity for a wider class of velocity fields is
undertaken in \cite{BPS04}. 

Summarizing, we note the following specific areas where future work would be of interest: 

\begin{itemize}

\item  the extension to time dependent velocity fields $v(x,t)$, either periodic in
time or random in time -- for example with an Ornstein--Uhlenbeck structure as in
\cite{paper1_stuart, inertial_1, inertial_2};

\item the extension to random velocity fields in space;

\item rigorous analysis of the parametric dependence of the effective
diffusivity on $\sigma$, $\tau$ and $\delta$, taking into account the
free streamline topologies;

\item further numerical studies for velocity fields other than the simple Taylor
Green flows studied here -- in particular to study problems where the Lagrangian
particle paths have open streamline topologies, such as the Childress-Soward family;
this is initiated in \cite{BPS04}.
\end{itemize}
\appendix
\section{Appendix}
\label{sec:app}
In this appendix we prove the existence of a unique invariant measure for the
process $\{x,\dot{x}\}$ solving \eqref{eqn:langevin} and, moreover, that the
generator $\LL$ of the process satisfies the Fredholm alternative. This
justifies the formal multi--scale
calculations presented in section \ref{sec:homog}. To simplify the notation we set
$\sigma = \tau =1$. The equations of motion become
\begin{equation}
\ddot{x} = v(x) -\dot{x} + \dot{\beta}.
\label{e:app}
\end{equation}
The generator of the Markov process $\{x,y \}$, with $y= \dot{x}$ is
$$
\LL = y \cdot \nabla_x + \left(v(x) -y \right) \cdot \nabla_y + \frac{1}{2} \Delta_y
$$
We have the following theorem
\begin{theorem}
\label{thm:ergodic}
Assume that $v(x) \in C^{\infty}(\T^d)$. Then there exists a unique, smooth
invariant density $\rho(x,y)$ for the process $\{x,y \}$:
$$
\LL^* \rho(x,y) = 0.
$$
Let further $h(x,y)$ be a smooth function such that $\int_{\R^d}\int_{T^d} h(x,y)
\rho(x,y) \, dx dy =0$. Then the Poisson equation
\begin{equation}
-\LL f = h
\label{e:pois_app}
\end{equation}
has a unique mean zero solution  in $L^2(\T^d \times \R^d, e^{-\delta^2 \|y \|^2}
dxdy)$  for every $\delta \in (0, 2 \sigma^{-2})$.
\end{theorem}
The proof of the existence of a unique invariant measure for our process is based on
the results of \cite{MatSt02} and is broken into three lemmas. First, we need
to prove the existence of a smooth transition probability density for our Markov
process. This is accomplished by means of H\"{o}rmander's theorem
\cite[Thm V38.16]{rogers2}. Then we need to prove the
compactness of phase space, for which we need to find an appropriate Lyapunov
functions. Finally, we need to prove that the transition probability density is
everywhere positive. To show this we need to use a controllability argument.
The proof
of the existence and uniqueness of solutions of the Poisson equation
\eqref{e:pois_app} is based on Fredholm's theorem.
\begin{lemma}
\label{lem:hypoell}
The Markov process generated by $\LL$ has a smooth transition probability density.
\end{lemma}
\proof This follows by an application of H\"{o}rmander's theorem. The basic idea
behind this theorem is that, even though noise does not act directly to the position
variable, there is nevertheless sufficient interaction between momentum and position
so that noise, and consequently smoothness, is transmitted to all degrees of
freedom. We write the generator in H\"{o}rmander's "sum of squares" form:
$$
\LL = \frac{1}{2} \sum_{i=1}^d X_i^2 + X_0,
$$
where $X_i = \frac{\partial}{\partial y_i}, \, i=1, \dots d$ and $X_0 = y \cdot
\nabla_x + \left(v(x) -y \right)$. Let now $[A,B]$ denote the commutator between the
vector fields $A, \, B$ and let $\mbox{Lie} \{F \}$ denote the Lie algebra generated
by the family of vector fields $F$. Define
$$
\mathcal{A}_0 = \mbox{Lie} \{ X_1, \dots X_d \}
$$
and
$$
\mathcal{A}_k = \mbox{Lie} \{[X_0, U], \,  U \in \mathcal{A}_{k-1} \}, \, k=1,2,
\dots
$$
Set finally
$$
\mathcal{H} = \mbox{Lie} \{ \mathcal{A}_0, \mathcal{A}_1 \dots \}.
$$
According to H\"{o}rmander's theorem, a sufficient condition for the Markov process
$\{x,y \}$ to possess a smooth invariant density is for $\mathcal{H}$ to span the
tangent space $T_{x,y}M$, where $M = \T^d \times \R^d$. We readily check now that
$$
[X_0, X_i] = - \frac{\partial }{\partial x_i} + \frac{\partial}{\partial y_i}, \quad
i=1, \dots d,
$$
and consequently
$$
\mbox{Span} \left( \mbox{Lie} \left\{\mathcal{A}_0, \mathcal{A}_1 \right\} \right) =
T_{x,y} M.
$$
Thus, H\"{o}rmander's hypothesis is satisfied and the Markov process generated by
$\LL$ has a smooth density. \qed

Now we prove that the existence of a Lyapunov function.
\begin{lemma}
\label{lem:lyapunov}
There exists a constant $ \beta >0$ such that the function $V(x,y) = 1
+ \frac{1}{2} \|y \|^2$ satisfies
$$
\LL(V(x,y)) \leq -  \frac{1}{2} V(x,y) + \beta.
$$
\end{lemma}
\proof We have that $V(x,y)$ maps the state space onto $[1, \infty)$ and that
$\lim_{\| y\| \rightarrow \infty}V(x,y) = \infty$. Moreover we have
\begin{eqnarray}
\LL (V(x,y)) & = & v \cdot y - \|y \|^2 + \frac{d}{2}
 \nonumber \\ & \leq & - \frac{1}{2} \|y \|^2 + \left( \frac{d}{2} +
 \frac{1}{2} \|v \|^2 \right)
  \nonumber \\ & \leq & -  V(x,y) + \beta,
\end{eqnarray}
with $\beta = d/2 + (1/2) \sup_{x \in \T^d} \|v(x) \|^2 + 1$. \qed

The last ingredient which is needed for the proof of the ergodicity of the process
generated by $\LL$  is
the fact the transition probability  $P_t$ is everywhere positive.
\begin{lemma}
\label{lem:control}
For all $z:=( x, y ) \in \T^d \times \R^d, t>0$ and open $\mathcal{O} \subset \T^d
\times \R^d $, the transition kernel for \eqref{e:app} satisfies $P_t(z, \mathcal{O})
> 0 $.
\end{lemma}
For the proof of this lemma we refer to \cite[Lemma 3.4]{MatSt02}.

{\it Proof of Theorem \ref{thm:ergodic}.} The existence of a unique invariant measure
 follows from Lemmas \ref{lem:hypoell}, \ref{lem:lyapunov} and \ref{lem:control},
upon using Corollary 2.8 from \cite{MatSt02}. In order to prove the existence and
uniqueness of solutions of the Poisson equation \eqref{e:pois_app}
we need to prove that the generator $\LL$ has compact resolvent. This is accomplished
in \cite[Thm 3.1, Thm 3.2]{per_hom_hypoell}. \qed
\begin{remark}
The above lemmas enable us to conclude that the system converges exponentially fast
to its invariant distribution \cite{MatSt02}.
\end{remark}
\begin{remark}
We a bit of extra work we can also prove sharp estimates for the invariant
distribution and the solution of the Poisson equation $-\LL f = h$. We refer to
\cite[Thm 3.1, Thm 3.2]{per_hom_hypoell} for details.
\end{remark}
\def\cprime{$'$} \def\cprime{$'$} \def\cprime{$'$}
  \def\Rom#1{\uppercase\expandafter{\romannumeral #1}}\def\u#1{{\accent"15
  #1}}\def\Rom#1{\uppercase\expandafter{\romannumeral #1}}\def\u#1{{\accent"15
  #1}}

\end{document}